# Genomics as a Service:
# a Joint Computing and Networking Perspective

G. Reali[1], M. Femminella[1,4], E. Nunzi[2], and D. Valocchi[3]

[1]Dept. of Engineering, University of Perugia, Perugia, Italy
[2]Dept. of Experimental Medicine, University of Perugia, Perugia, Italy
[3]Dept. of Electrical and Electronic Engineering, UCL, London, United Kingdom
[4]Consorzio Interuniversitario per le Telecomunicazioni (CNIT), Italy

*Abstract*—**This paper shows a global picture of the deployment of networked processing services for genomic data sets. Many current research and medical activities make an extensive use of genomic data, which are massive and rapidly increasing over time. They are typically stored in remote databases, accessible by using Internet connections. For this reason, the quality of the available network services could be a significant issue for effectively handling genomic data through networks. A first contribution of this paper consists in identifying the still unexploited features of genomic data that could allow optimizing their networked management. The second and main contribution is a methodological classification of computing and networking alternatives, which can be used to deploy what we call the Genomics-as-a-Service (GaaS) paradigm. In more detail, we analyze the main genomic processing applications, and classify both the computing alternatives to run genomics workflows, in either a local machine or a distributed cloud environment, and the main software technologies available to develop genomic processing services. Since an analysis encompassing only the computing aspects would provide only a partial view of the issues for deploying GaaS systems, we present also the main networking technologies that are available to efficiently support a GaaS solution. We first focus on existing service platforms, and analyze them in terms of service features, such as scalability, flexibility, and efficiency. Then, we present a taxonomy for both wide area and datacenter network technologies that may fit the GaaS requirements. It emerges that virtualization, both in computing and networking, is the key for a successful large-scale exploitation of genomic data, by pushing ahead the adoption of the GaaS paradigm. Finally, the paper illustrates a short and long-term vision on future research challenges in the field.**

## I. INTRODUCTION

THIS paper gives a comprehensive description of the ongoing initiatives aiming at increasing the usability and effectiveness of genomic computing by leveraging networking technologies. The motivations that have stimulated a fruitful trait-union between the genomics and networking essentially are represented by the need of supporting the modern medical activities making an extensive use of a massive and rapidly increasing genomic data, stored in repositories accessible through the Internet. These activities have been established over the last fifteen years, since the successful completion of the Human Genome project, in 2003, which required years of intense research. At that time, although the importance of results was clear, the possibility of handling the human genome as a commodity was far from imagination due to costs and complexity of sequencing and analyzing complex genomes. Today the situation is different. The progress of DNA (deoxyribonucleic acid) sequencing technologies has reduced the cost of sequencing a human genome, down to the order of 1000 € [2]. Since the decrease of these costs is faster that the Moore's law [4], two main consequences are expected. First, it is easy to predict that in few years a lot of applicative and societal fields, including academia, business, and public health (e.g., see [6]), will make an intensive use of the information present in DNA sequences. For this purpose, it is necessary to leverage interdisciplinary expertise from different disciplines, including biological science, medical research, and information and communication technology (ICT), which embraces data networking, software engineering, storage and database technologies, and bioinformatics. The impact of this process is significant in many application areas such as medicine, food industry, environmental monitoring, and others. The execution of genomic analyses requires significant efforts in terms of manpower and computing resources. Unfortunately, the cost of setting up large computing clusters and grids to efficiently perform such data analyses can be afforded by just few specialized research centers [14]. Since it cannot be assumed that any potential user owns the infrastructure for massive genome analysis, a cloud approach has been envisaged [19][5].

The second consequence is that, under a practical viewpoint, the cost to produce a unit of genomic data decreases more rapidly than the cost for storing the same unit and distributing it. Thus, given this trend, the bottleneck for handling genomics data will reside on the ICT side [5]. In other words, the most critical element of a networked genomic service is not the sequencing capability of machines, but the capacity of processing large data sets efficiently due to the limitations of accessing and exchanging data remotely. In fact, it is expected that genomics will be more demanding than astronomy, YouTube, and Twitter in terms of data acquisition, storage, distribution, and analysis [7]. The urgency of finding suitable networked solutions for managing such a huge amount of data is also witnessed by the fact that the Beijing Genomic Institute is compelled to ship hard drives for delivering genomic data

[15][25].

Genomic data management can be classified as a Big Data problem [9][10], according to the classical 3V (Volume, Velocity and Variety) model [54]. The size of a single human raw genome is roughly 3.2 GB and the global production rate is increasing over time with an exponential growth rate. Moreover, the bioinformatic processing tools, which are typically organized in software pipelines, make large use of metadata having a total volume sometimes even larger than raw data. Even these metadata, retrievable from reference databases, have to be distributed through the available networks for implementing networked genomic services [15]. The suitable handling of genomic data sets requires re-considering some aspects of data management already developed for managing other data types, such as the content growth rate, the content popularity variations over time, and the mutual relationships between genomic data. These aspects are illustrated in this paper, along with the relevant data management solutions.

These three aspects, namely the need to (i) resort to a cloud computing model for processing genomic data sets, (ii) design new networked solutions for accessing and exchanging huge amount of genomic data, and (iii) design novel data management policies to address their specific features, all together contribute to the definition of Genomics-as-a-Service (GaaS). Thus, GaaS is a novel paradigm that is rapidly gaining ground for processing genomic data sets based on the cloud computing technologies. It includes not only networking aspects, which could be either a bottleneck or a flywheel for a widespread usage of genomics in multiple fields, but also the specific features of datasets and their usage. The latter aspect could both generate significant issues and offer great exploitation potentials for network and service management.

To sum up, the main contributions of this paper are:
- Illustrating the technical problems and the still unexploited features that could allow optimizing the networked management of genomic data. In particular, the aim is to overcome or integrate the typical solutions already used to manage other types of big data, in order to improve the effectiveness of use of the GaaS instances.
- Giving an overview of widely used genomic processing applications, for medical and research activities, with a particular emphasis on open-source components and their impact on the network resource management, including a critical evaluation of the computing alternatives for GaaS implementation.
- Presenting the main ongoing activities related to the networked management of genomic data, together with a discussion on the most suitable networking alternatives for GasS deployment.
- Giving both short and long-term visions on future research challenges in the field, with a special emphasis on computing and networking issues and potential future implementation venues of GaaS in the upcoming fifth generation mobile services (5G) service architectures [238].

The structure of the paper is as follows. In section II, we give a comprehensive view of the background, emphasizing the use of genomes and related challenges. In section III, we present the related works in the field and review other surveys in the genomics and Big Data applied to medicine, highlighting the original contributions of this paper. In section IV, we present the peculiarities of genome content management and their potential impact on optimization of network and data management policies. The subsequent section V focuses on genomic computing alternatives for GaaS systems. In particular, it deals with genomic applications and tools used for genomics processing. These findings are summed up in two taxonomies for genomics computing, one about computing infrastructures used for genomics computing, and the other about software technologies for implementing genomics pipelines. Finally, we also present two specific genomics processing case studies, analyzed to show main peculiarities of two real genomic pipelines, highlighting computing and networking requirements. Section VI mainly focuses on classifying networking approach to support GaaS. In this regard, we present two taxonomies, one relevant to wide area network techniques, and another to datacenter networking. For each technique, we discuss pros and cons in the light of the application framework and ease of usage. Section VII describes open research challenges, with emphasis on the aspects related to networking, computing, and privacy, both in the short and long term. Finally, Section VIII draws some final considerations.

## II. BACKGROUND

DNA and RNA (ribonucleic acid) are macromolecules that store the genetic information of any living body. They have a periodic helicoidal structure, which is analyzed by biologists for extracting information related to multiple aspects of life, including growth, reproduction, health, food production, evolution of species, more recently even for exploiting these molecules as a medium for storing information [52], and many others.

In more details, the DNA, is formed by two strands of nucleotides, or bases, commonly indicated by using the initial letter of their name: A (adenosine), C (cytosine), G (guanine) and T (thymine). Subsequent nucleotides in each strand are joint by covalent bonds while nucleotides of two separated strands are bound together with hydrogen bonds thus making the double DNA strand. The identification of significant combination of these bases, commonly referred to as *genes*, and their mutual relation (*genotypes*), is the research focus of genomic scientists, which are still struggling to associate them with any macroscopic features of bodies (*phenotypes*). The overall sequence of nucleotides encodes roughly 27,000 genes and is organized in 23 chromosomes. This research field is still in its early stage, since most of the genetic information stored within DNA is still unknown [53], even if the mere binary size of a human DNA is about 3.2 GB.

| Sequencer | 454 GS FLX | HiSeq 2000 | SOLiDv4 | Sanger 3730xl |
|---|---|---|---|---|
| Sequencing mechanism | Pyrosequencing | Sequencing by synthesis | Ligation and two-base coding | Dideoxy chain temrmination |
| Accuracy (%) | 99.9 | 98 | 99.94 | 99.999 |
| Output data/run | 0.7 Gb | 600 Gb | 120 Gb | 1.9-84 kb |
| Time/run | 24 Hours | 3-10 Days | 7-14 Days | 20 Mins-3 Hours |
| CPU | 2 Intel Xeon X5675 | 2 Intel Xeon X5560 | 8 2.0 GHz processors | Pentium IV 3.0 GHz |
| Hard Disk size | 1.1 TB | 3 TB | 10 TB | 280 GB |

## A. DNA sequencing

The increasing usage of genomes has been eased by the technical progresses of sequencing machines since the sequencing costs decreased more quickly than the Moore's Law during the last 15 years [1].

A comprehensive survey and comparison of modern sequencing techniques, referred to as Next Generation Sequencing (NGS) techniques, can be found in [47]. Different sequencers can offer different performance in terms of sequencing time, accuracy of results, output size, throughput, due to different sequencing mechanisms and hardware configurations. Table 1 reports a summary comparison of different sequencers in terms of sequencing mechanisms, expected performance, and hardware configuration [47].

## B. The use of genomes and related challenges

Although the expectations of genomic research extend well beyond the known results, the current achievements have already reshaped a lot of human activities and, clearly, this impact is believed to dramatically increase in the next few years. For example, in pediatrics the usage of genomics allows the early and accurate prognosis, management, surveillance and genetic advice of some rare diseases [232] and particularly aggressive cancers, such as neuroblastoma [233]. In addition to medicine, other fields of human activities have witnessed significant benefits due to the introduction of genomic assisted techniques. For instance, in forensic science genomic analysis can be used for providing a scientifically defensible approach to questions of shared identity [235]. In agriculture, genomic-assisted plant breeding strategies are used to develop plants in which both crop productivity and stress tolerance are enhanced [231]. In animal breeding, the results in genomics have led to an extensive application of genomic or whole-genome selection in dairy cattle [230]. In food production, the segment of food processing aids, i.e. industrial enzymes enhanced by the use of genomics and biotechnology, has proven invaluable in the production of enzymes with greater purity and flexibility, while ensuring a sustainable and cheap supply [234].

Research, medical, and business-related activities make extensive use of bioinformatics software packages and databases. The information data typically used for genomic processing are included not only sequenced genomes (e.g. 1000 genomes database [100]). In fact, it typically requires additional *auxiliary files*, which could include known DNA sequences stored in dedicated databases (e.g. GenBank [11], UniProtKB/Swiss-Prot [12], Protein Information Resource, PIR, [13]), genomic models (e.g. GRCh38 human genome model available in Genome Reference Consortium databases [101]), and/or mutual relationships of genetic patterns (e.g. human disease network [49][50]). Thus, any genomic analysis requires the combined usage of different files representing the sequenced genomes and auxiliary files. Consequently, each processing step may require an amount of data ranging from tens to hundreds GB, depending on the target bioinformatic analysis. In addition, genomic processing may require repeated executions as in comparative studies. Retrieval, management, processing, and storage of this huge amount of data poses enormous challenges not only to computer engineers, but also to networking researchers, since all network resource categories are highly involved, including storage space, processing capacity, and network bandwidth. The related technical issues are expected to become increasingly challenging since, for example, is the near future all newborn will be sequenced, and most of the future medicine (P6 Medicine: Personalized, Predictive, Preventive, Participatory, Psychocognitive, and Public) will be based on genomic computing [46].

## III. RELATED WORK

This work updates and integrates a number of survey papers, dealing with genomics and the relevant hardware/software tools, with the relevant networking aspects.

One of the first work dealing with the possibility to run genomics services in cloud is [185]. The authors analyze the trend of the cost of DNA sequencing, and consider the possibility of migrating the genomic processing in the cloud, from a high-level perspective. A more in-depth analysis is carried out in [19], which deals with computing infrastructure to support genomics services. It tries to provide a comparative view about the possibility of running genomic processing on clusters, grid, cloud, and heterogeneous acceleration hardware. In addition, it provides a first outlook in the usage of the MapReduce paradigm in genomics. This work is extended by [9], which focuses on the Big Data nature of genomics, and on the possibility to process them in cloud by means of the three canonical paradigms: infrastructure-as-a-service (IaaS), platform-as-a-service (PaaS), and software-as-a-service (SaaS). In addition, it analyzes the possibility to use Hadoop, an open source implementation of the MapReduce paradigm. It also lists a number of packages that already implement the MapReduce paradigm.

The survey [116] generically indicates genomics as one of the possible Big Data applications in bioinformatics, and provides a general outlook on the different computing paradigms that could be suitable for this scope (again, cluster, grid, cloud, hardware acceleration). It also illustrates some issues relevant the heterogeneity of semantics, ontologies, and open data format for their integration, which is one of the main issues in the deployment of genomics technologies in cloud networks.

The recent survey [5] deals with the execution of genomic analysis services in high performance computing (HPC) environments and discusses their deployment in cloud resources. In addition, they provide a throughout description of the development of a SaaS approach, used to simplify the access to HPC cloud services for carrying out mammalian genomic analysis.

The paper [225] provides an economic analysis relevant to the use of cloud genomic services by individual scientists, and recommend funding agencies to buy storage space for handling genomic data sets in the most popular cloud services. In this way, authorized scientists would be able to easily and cheaply access a *global commons* resource whenever they need, without wasting money in buying individual access to cloud services for genomics.

The comprehensive survey and evaluation of NGS tools [180] provides a valuable guideline for scientist working on Mendelian disorders, complex diseases and cancers. The authors surveyed 205 tools for whole-genome/whole-exome sequencing data analysis supporting five analytical steps: (i) quality assessment, (ii) alignment, (iii) variant identification, (iv)variant annotation, and (v) visualization. For each tool, they provide an overview of the functionality, features and specific requirements. In addition, they selected 32 programs for variant identification, variant annotation and visualization, which were evaluated by using four genomic data sets.

The survey [229] has a different focus. It reviews several state-of-the-art high-throughput methodologies, representative projects, available databases, and bioinformatic tools at different molecular levels. They are analyzed in the context of the different areas, as genomics and genetics variants, transcriptomics, proteomics, interactomics and epigenomics.

Finally, a classification of design approaches of several pipelines is provided in [187], together with the description of some specific applications in important research centers. They were further commented in the more recent survey [186]. This last work focuses on more modern approaches than traditional ones, based on scripting and makefiles, and provides useful indications based on analysis requirements and the user expertise.

Finally, some papers deal with aspects related to privacy and security. In particular, [226] focuses on all aspects of security in clouds, including requirements for cloud providers, encryption techniques and need of suitably training personnel. The paper [227] discusses potential issues specifically related to privacy. They characterize the genome privacy problem and review the state-of-the-art of privacy attacks on genomic data and the relevant strategies to mitigate such attacks, by contextualizing them in the perspective of medicine and public policy. Finally, they present a framework to systematize the analysis of threats and the design of countermeasures in a future perspective.

The original contribution of our work with respect to the above mentioned papers is essentially twofold. First, it includes a comprehensive classification of both the platforms, which are not limited to clouds, suitable to run genomic software pipelines, and the software technologies used to develop such pipelines. Second, to the best of our knowledge, this is the first work that jointly classifies both computing aspects related to genomics and the relevant networking issues, which will have a significant impact on the adoption of cloud genomics in the near future.

## IV. THE UNEXPLORED FEATURES OF GENOMIC DATA SETS

From the point of view of a networking scientist, the management and exchange of genomic files could heavily affect performance of the whole networked system if their management tools and strategies are those used for other Big Data types, and the genomic data peculiarities are ignored. In fact, the usage and the features of genomic files differ substantially from those of generic internet contents. Without appraising the existing differences, the resulting network and content management techniques would be highly suboptimal. Hence, in order to propose criteria for optimizing the network performance for supporting transfer and processing of genomic data, it results of critical importance to emphasize the original features related to genomic contents in comparison to other contents, typically available on-line. These features can be organized in the following categories: *content growth rate*, *content popularity*, and *logical relationships between genomic contents*. These features can both generate significant issues in the management of genomic data sets and, considered all together, offer a great potential for designing ad hoc data management policies for the GaaS paradigm. In fact, if it is true that massive amounts of data represent always a challenge for ICT infrastructures, it is also true that discovering logical relationships between contents may allow forecasting their upcoming popularity. In turn, this popularity can be exploited by means of content replication policies to ease the access to genomic contents. Finally, we also analyze typical big data issues in the context of these concepts.

### A. Content growth rate

The growth of genomic data is unprecedented. In [48] Schatz and Langmead say that "The roughly 2000 sequencing instruments in labs and hospitals around the world can collectively sequence 15 quadrillion nucleotides per year, which equals about 15 petabytes of compressed genetic data". The generation of this amount of data is a tremendous challenge for all network management activities. For example, the design of a suitable data storage and retrieval system must cope with the following issues:

- *Access transparency*: make data accessible regardless the user locations.
- *Location transparency*: make data accessible after any change of the repository locations.
- *Availability*: according to the CAP theorem [42], a distributed information system cannot guarantee consistency, availability, and partition-tolerance at the same time. The suitable trade-off has to cope with both storage issues and the tolerable service time, along with the metrics illustrated below.

- *Failure transparency or Partition tolerance*: data provisioning must be robust to link and router failures. This metric is strictly related to access transparency.
- *Consistency*: storage and cache instantiation and update procedures must guarantee data and metadata consistency. This metric is strictly related to location transparency.
- *Scalability*: the effort for managing any increase of the network load must scale gracefully. Even scalability has to be optimized in relation to the suitable trade-off illustrated by the CAP theorem.

In this regard, the recent paper [7] defines genomics as a "four-headed beast", considering the computational demands across the 4-phase lifecycle of a dataset: acquisition, storage, distribution, and analysis.

In genomics, data acquisition is highly distributed and involves heterogeneous formats and production rates. The sequencing centers range from small laboratories, with a few instruments generating a few terabases per year, to large dedicated facilities, which can produce several petabases a year [7][48]. This heterogeneity is one of the distinctive characteristics of genomics, which should be taken into account when it is necessary to move contents from the sequencing centers to somewhere else.

A more aggressive estimate forecasts a world's population close to 8 billion by 2025, with sequenced genomes of about 25% of the population in developed nations and half of that in less-developed nations. This growth rate exceeds by far the other three domains producing Big Data (astronomy, YouTube, and Twitter) [7].

In addition, new single-cell genome sequencing technologies are revealing unknown levels of variations, especially in cancers, which call for sequencing the genomes of thousands of separate cells within a single tumor [43]. Finally, when other applications and research areas making use of sequenced DNA enter the game and require the analysis of transcriptome, epigenome, proteome, metabolome, and microbiome sequencing (i.e. all the '-omics'), they can require sequencing the genetic material multiple times per subject so to monitor molecular activity, which further increases content growth rate [176].

Summing up, content growth rate is a challenge for big data platforms implementing GaaS, not only for managing data repositories, but also for the distributed production footprint. The potential impact is not only on storage facilities, but also, and especially, for transport services, which could be requested to move unprecedented volume of data. At the same time, digging into the contents and finding out novel relationships between data sets may help not only to manage the challenge, but also to take advantage of it. The next two subsections, dealing with estimation of content popularity and logical relationships between contents, explore these concepts.

### B. Content popularity and secondary use

In general, data processing time is affected by multiple factors, including the computing capabilities of nodes/clusters and the time needed to move contents where they have to be elaborated. In this regard, content distribution solutions play a major role, since they are able to decrease transfer time by increasing content locality through replication [175]. Distributed storage and caching in networks can be highly optimized by considering content popularity, in order to make data easily available from where and when it is assumed they will be requested. In recent years, some studies have been done to identify the most common patterns in content popularity and most suitable models [168]. For typical Internet data types, such as YouTube video clips, popularity typically increases over time until a maximum is reached and then it decreases [168], with some variations which can lead to spurious or periodic rebounds [169]. A general belief is that predictions are possible due to the regularity with which user attention focuses on content. In particular, predictions typically result more accurate for contents whose popularity fades quickly, whereas those for content with a longer life cycle (e.g. video clips) are prone to errors [170].

The popularity evolution of genomic contents is not known in depth. A genome is a really different content from those typically available on the web. It is a plentiful source of information, most of which is still unknown [171]. It may happen that the interest over a particular genomics dataset is low for a long time (even years), and then begins to increase due to new research achievements and the need of re-investigating some genome properties, even potentially different from those for which is was initially collected and made available, the so-called secondary use.

Thus, the evolution of the popularity of genomic contents can evolve over time in a still unpredictable manner. As a consequence, in the short and medium terms, beyond the urgent need to have large amounts of storage capacity, specific investigations for managing data storage (caching and replication, [168]) are still expected and could significantly contribute to optimization of genomic content distribution. Clearly, analyses able to discover hidden relationships between different datasets could allow anticipating the popularity variations of some of them, thus facilitating the management of geographically distributed replicas.

However, genomic contents are not simple files that can be freely distributed on the web, since they include personal information. In particular, when dealing with the processing of biological samples, things are never easy. The two main issues to be considered for users' protection are privacy, which should be ensured by anonymization procedures (or more precisely, de-identification), and autonomy, which regards the potential secondary use of users' samples, governed by informed consent. Thus, caching or replication of genomics datasets seems feasible only for anonymized contents for which informed consent has been issued. We now briefly discuss the anonymization, secondary usage, and informed consent for digital genomic data sets.

Content anonymization has pros and cons [174]. The obvious advantage is that it provides some privacy protection, which may be mandatory for handling human samples. Said this, it presents also significant drawbacks. In fact, it may put at risk the scientific value of the biological samples, as anonymous data are more difficult to check and validate. In addition, and even more important, the lack of direct connection between biological samples and donors could make it difficult to track changes of patients' condition over time. In addition to these cons, anonymization itself is difficult to guarantee, since the genomic information is unique, and it has embedded identifying characteristics. Furthermore, companies offering sequencing services may have loose policies in handling customer privacies. In this regard, in [174] the authors have analyzed the risk of re-identification, by comparing the sequencing services offered by four companies [228]. Their analysis reveals that information provided by companies to consumers is neither clear nor complete, especially about the risk of re-identification, thus undermining the validity of consent. Thus, the analysis indicates that companies providing sequencing services should improve the transparency regarding their handling of consumers' samples and data, including an explicit and clear consent process for research activities.

As for the secondary use of de-identified data, the study presented in [173] examines the potential negative consequences of limited oversight on available genomic datasets. This analysis reveals that the risks of misuse of available datasets cannot be completely eliminated by anonymizing individual data. To this aim, the authors suggest setting up a Data Access Committee to review proposed secondary uses. This committee should be a mandatory component of the trustworthy governance of any repository of data or biological samples.

Although informed consent should protect autonomy of patients/donors from unwanted secondary use, the real issue is that biobanking with prospective consent is only a relatively recent movement, and these biobanks often do not contain enough samples for specific analyses, such as those concerning rare diseases or specific populations. On the other hand, accessing archival sample obtained without specific consent for secondary usage can be not only expensive, but also illegal, even if these samples can be really valuable for research. Although the authors in [174] present a procedure for waiver of consent in some specific cases, it is clear the problem is still open.

Summing up, it is clear that although anonymization and prospective informed consent of personal genomic data are still in their infancy, they are needed procedures for handling human samples. However, in order to avoid them to hinder the possibility to exploit content popularity and logical relationships between genomic contents, novel robust and flexible policies still need to be defined.

### C. Ontologies and semantics

As it happens in other fields producing Big Data, the increasing amount of genomic data poses serious challenges on their organization aimed to help the extraction of the embedded information. In this regard, genome annotation is defined as the process of identifying the locations of genes, the coding regions in a genome, and their associated functions. Thus, once a genome is sequenced, it needs to be annotated.

Some initiatives have been started for addressing this issue, such as Gene Ontology Annotation (GOA) [117] project, which aims at integrating the protein information accessible through the UniProt database with other databases. The project, promoted by the Gene Ontology (GO) consortium, makes use of a wide dynamic vocabulary of several thousands of terms used to describe molecular functions and protein features. The interested reader can find more details on GOA and similar initiatives in [118].

A closely related aspect, which can have a significant impact not only on data integration but also in their distributed storage and networked access in GaaS platforms, consists in the logical data relationships over a further dimension, which indicates the set of genes related to a disease or any other macroscopic biologic features, commonly referred to as phenotypes. It is known that the exploitation of logical relationships between data can help define content management strategies. For what concerns genomic data, particular relationships can be found, and exploited, between genes (nucleotide patters) that are shared by different phenotypes. An example can be found in Fig. 1, which shows a portion of the genetic relationships of human diseases, mapped on the basis of the findings in [49] and made available by [51] under the name "Diseasome". In Fig. 1, circles indicate diseases, the circle size increases with number of genes characterizing a disease, and arcs connecting two circles mean a shared genetic content. For example, Fig. 1 shows the entire network of genetic diseases; the zoomed part highlights that colon cancer and leukemia share a significant amount of genetic contents. Hence, in case a colon cancer diagnosis is investigated, it can be assumed that other connected diseases should be investigated as well.

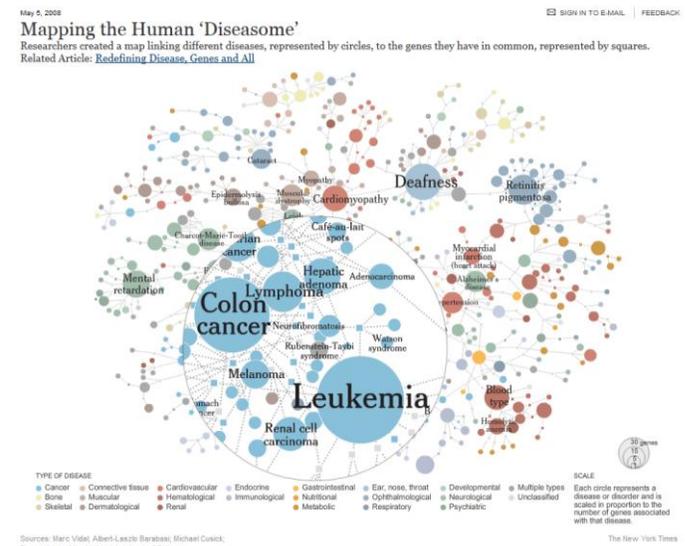

Fig. 1. Graph representation of human "Diseasome", with a zoom on connections between leukemia and colon cancer genes [51].

In this regard, the adoption of machine learning and artificial intelligence techniques for genomics analysis [183] can provide novel and significant insights.

Machine learning techniques may be useful especially in the direction of improving predictive models, which can help to put in close relationship apparently uncorrelated diseases. The discovery of these relationships may help setting up specific data management policies in GaaS platforms. For instance, some popular data might be pre-cached in suitable positions for fulfilling future requests.

In section V.B we discuss potentials and limitations of current machine learning techniques applied to genomics.

### D. Genomic Big Data issues

Classification of the genomic data management problem as a "Big Data" one seems to be quite obvious. Nevertheless, we formalize this classification by using the Gartner's 3V model [54] and expectations [57]. In [54], *"Big Data"* is described as *"high-volume, high-velocity and high-variety information assets that demand cost-effective, innovative forms of information processing for enhanced insight and decision making"*.

In terms of data *volume*, the expected diffusion of genomic data in the next future is indeed characterized by overwhelming volumes of data that require suitable management, as already analyzed in section IV.A. This management requires to dig into aspects still under-investigated, such as genomic indexing, redundancy management, popularity and retention. V*elocity* is a service requirement. For example, it could be determined by the medical needs of handling serious diseases in a short time. From the ICT perspective, it requires suitable storage policies of operational data, dynamic caching, and suitable trade-off between latency and ICT resource exploitation. For what concerns *variety*, the combined usage of different data types, such as genome files, alignment files, genomic annotations, reference genome models, software tools, and related signaling generates different challenges. One of them is generated by the production and sharing of metadata and their management. In fact, being genomics a relatively recent discipline, the individual development of software packages and data structure has produced a tower of Babel of metadata structures indicating the processing output. A lack of standardized metadata representation is a potential huge problem that requires significant actions. In particular, sequencing machine precision and reliability of results illustrated in statistical terms are essential aspects over which a common representation and understanding is necessary, not only for interpreting results, but also for retrieving metadata from distributed storage systems through distributed query management.

As mentioned in [57], "'Big Data' Is Only the Beginning of Extreme Information Management", involving multiple dimensions in management strategies, and genomics is prone to increasing data management complexity due to its effects in different fields, from research and medicine to multiple business areas.

## V. Genomic Computing

Nature.com defines genomic analysis as "*the identification, measurement or comparison of genomic features such as DNA sequence, structural variation, gene expression, or regulatory and functional element annotation at a genomic scale. Methods for genomic analysis typically require high-throughput*

*sequencing or microarray hybridization and bioinformatics*" [3]. Given a so ample definition, it is clear that the different types of genomic analysis, and thus relevant computing counterparts, are really so many and continuously increasing. Therefore, it is very difficult to find a single work encompassing all of them.

The -omics (whole-genome, whole-exome, transcriptome) data processing is typically performed through a pipeline of different software packages. The general theory of the implementation of these pipelines is beyond the scope of this paper since a wide and rapidly evolving scientific literature is available based on the target bioinformatic analysis [20][114]. In the following subsections we describe some basic, well-known genomic computing applications, followed by a list of genomic processing tools, which can be used to carry out most of these bioinformatics analyses. The subsection V.C shows two taxonomies, which categorize not only the computing infrastructure for genomic processing, but also the way the genomic processing software is typically organized (i.e. processing pipelines). These concepts are essential to give the reader a global view of the numerous computing alternatives available to implement GaaS instances, by highlighting pros and cons, especially in terms of flexibility, computing efficiency, and ease of usage.

Finally, we present the experimental analysis of two specific processing genomics pipelines, used to extract some insights and requirements to be used in the following Section VI, dedicated to the networking aspects.

### A. Genomic computing applications

The research on bioinformatics over the last decade has been shaped by the need of finding solutions to some challenging computing problems related to genomics. In this section, we illustrate some significant examples.

#### 1) Prediction of Protein Coding Regions in DNA sequences

This is essentially a *pattern-matching* problem. It consists in finding the portions of DNA strands including particular sequences of nucleotides [58][70]. This is the essential step of gene annotation and creation of metafiles [20]. The algorithms designed for solving this problem are based on searching short range correlations in the nucleotide arrangement though Discrete Fourier transform (DFT). DFT is computed over indicator binary vectors for each nucleotide. Each position of the vectors maps the presence or absence of a nucleotide with a binary 1 or 0, respectively [58]. However, the portions of DNA encoding production of proteins is only a minimal part of the whole available information. As discussed by the ENCODE project, the remaining part seems to be used by nature to encode functional aspects [53].

#### 2) Clustering Microarray Data

Microarray is a widely used technology for gene expression analysis. A microarray is a matrix of thousands of elements, each associated with a functional sequence of nucleotides. Since it allows determining the number of genes expressed in different conditions, it has become a widely used tool for research and diagnostic activities. For example, through microarray analysis it is possible to determine the influence of a gene in causing a disease. Since the amount of data produced is enormous, their clustering is essential for identifying functional group of genes [59]-[69]. Some specific genomic features must be considered in the design of clustering algorithms. For example, as mentioned above, a gene may be involved in different biological processes, such as a disease, thus, clustered regions could overlap. In addition, a gene expression may appear under different regulating conditions, and may change over time. Hence, clustering algorithm should consider time sequences and not mere static results of a single experiment.

### 3) Detection of single-nucleotide polymorphism (SNP)

It consists in finding any single-nucleotide variation in the DNA sequence [71]. It is the most common polymorphism, and it is believed to be related to different phenotypic features, such as disease susceptibility. Recent algorithmic approaches to the SNP problem make use of Bayesian methods [72].

### 4) Detection of Copy-Number Variation (CNV)

The number of copies of a particular gene in the genotype of an individual is referred to as *copy number*. They may span over large segments of DNA. It was found that, encompassing a significant number of genes, variations of this number had a significant role in evolution of species. For the same reason CNVs have important roles both in human disease and drug response. For this reason, the CNV analysis has been an intense genomic research area [73]-[79]. Some of the genomic processing experiments illustrated in what follows deal with CNV analysis, given its importance.

### 5) Differential Expression (DE)

The differential expression (DE) is an RNA analysis. It allows identifying the differentially expressed genes and/or transcripts in different experimental conditions. Even if at functional level it may be regarded as alternative to the use of microarrays, significant differences exist. For example, whilst the microarray analysis provides one measurement for each known gene, the sequenced RNA can be processed in order to find gene expression also in regions not previously annotated and to analyze multiple transcripts for individual genes. Nevertheless, it is worth to consider that the datasets of sequenced RNA (RNA-seq) to be analyzed are large and complex. This poses further challenges for suitable interpretation and for timely computing, which asks for high-throughput technologies providing results in acceptable times [80][81][82]. For this reason, one of the case studies for genomic processing and the relevant experiments illustrated in what follows is a DE analysis.

### 6) Sequence alignment

Comparison of gene or protein sequences of nucleotides can be accomplished if the overall sequences they belong to are suitably aligned [26]-[33]. In this way, their similarity, in terms of elementary bases or complex amino acid, can be analyzed. The likelihood of sequences is the outcome of the analysis, which is expressed as an alignment score, proportional to the number of matching position at each tentative alignment. A *global* alignment is the alignment which produces the largest score by using all characters from each sequence. A *local* alignment consists in finding the alignment that produces the largest score in local regions. In order to reduce the probability of determining false optimal alignments, the estimated probability of this event is used in conjunction with the alignment indicator.

Sequence alignment is involved in most of genomic processing illustrated above, and its role will be further analyzed in what follows, in regard to the computing requirements of the two experimental analysis presented in subsection V.D.

### B. Genomic processing tools

The computational needs of processing genomic data sets have stimulated the implementation of different genome processing tools. In this section, we mention some of the mostly used tools for the genomic computing purposes mentioned above.

We begin with alignment tools, given their central role in many genomic analyses. Basic Local Alignment Search Tool (BLAST) is a very popular DNA sequence alignment software tool and uses different alignment procedures depending on the sequence types (protein or nucleotide) of the query and on the database sequences. The algorithm core is based on a heuristic algorithm that approximates the Smith-Waterman algorithm [83]. Solutions for accelerating the execution of BLAST have recently been proposed. They include its parallel execution on shared memory HPC (SGI Altix [37]), distributed-memory HPC (IBM BlueGene/L [41]), and execution in clusters implemented by using high-speed interconnections (MPP2 [36]). Also, proposals for executing parallel BLAST methods in general clusters exist [34][35][40]. Typically, these approaches aim at speeding up the execution of BLAST queries by resorting to partitioned database structures. For instance, the mpiBLAST package [34] can achieve super-linear speed-up with the size of databases by removing unnecessary paging. However, some scalability issues [36] along with other problems due to merging results and I/O synchronization [40] still need to be managed carefully.

A different heuristic for generating gapped alignments is implemented in gapped-BLAST [27]. It is a program which is claimed to be much faster that BLAST. Position-Specific Iterated BLAST (PSI-BLAST) [26][27] has a processing speed similar to gapped BLAST, and can detect weaker, although biologically relevant, similarities. Similar functions are implemented in the FASTA algorithm [30], which is a DNA and protein sequence alignment software package, SSEARCH [28], which implements the Smith-Waterman algorithm [83] rigorously for determining the degree of similarity between a query sequence and a group of sequences of nucleotides acid or proteins. The rigorous implementation of the Smith-Waterman algorithm makes FASTA execution slower than BLAST.

IMPALA [32], is a software package designed for comparing a single query sequence with a database of position-specific score matrices generated by PSI-BLAST. Its sensitivity to biologically relevant similarities is similar to that of PSI-BLAST. It makes use of statistical significance of detected scores and implements the Smith–Waterman algorithm rigorously. CUSHAW [29] is a parallelized short read aligner. It makes use of the processing power of graphic adapters, based on the compute unified device architecture (CUDA). Its alignment algorithm is based on the Burrows-Wheeler transform (BWT). SAM-T98 [31] is a method for searching a target sequence. It iteratively builds a hidden Markov model (HMM) used for database search and finding remote *homologs* of protein sequences (homology between protein or DNA sequences is related to shared ancestry). HMMER [33] is a tool implemented by using HMM statistical methods for searching sequence databases for homologs of protein sequences, and aligning protein sequences.

The SNP search can be done by using different software packages, such as include PolyPhred [84], which compares fluorescence-based sequences across traces obtained from different individuals to determine single nucleotide variations. Similar fluorescence-based techniques are used by SNPdetector [85], and novoSNP [86].

Another family of SNP detectors are software tools that analyze reads generated by the next-generation sequencing (NGS) machines, by comparing sequence differences among DNA samples, such as MAQ [87], GATK [88], Atlas-SNP2 [71], SAMtools [89], and VarScan [90].

Two surveys of computational tools for *variant* analysis detection using NGS data are shown in [91] and [180]. In subsection V.D, we show a case study implementing the Mean Shift-Based (MSB) algorithm [92]. According to this algorithm, the adjacent data windows with similar read depths (i.e. the number of times a nucleotide is read) are merged together along chromosomes. If the read depths of a sliding window are significantly discordant with the depths of the merged windows, a breakpoint is reported. This algorithm is implemented in two software tools, CNVnator [93] and BIC-seq [94]. The former, used in a case study illustrated in what follows, makes use of single individual samples. It can detect CNVs of different size, ranging from hundreds of bases to megabases.

The need of analyzing high-throughput sequencing data has led to the implementation of HTSeq [105]. It is a package written in Python, which allows writing custom scripts for different types of analyses, such as statistical evaluation of the data quality, reading in annotation data from a GFF (General Feature Format) files, association of aligned reads from an RNA-Seq to exons, which are sequences of nucleotides encoded by a gene and present in the used strand of RNA. Among the available functions, *htseq-count* is used in the *differential* expression analysis for comparing the expression of genes in different samples. HTSeq can be used also in conjunction with DESeq, which is an R software package designed to analyze count data of RNA-seq and makes use of a refined model based on the negative binomial distribution to detect differential expression.

It is worth mentioning other software tools typically used in genomic computing. For example, FastQC is a tool written in Java, which is used for quality control on sequences produced by high throughput sequencing machines [95]. Trimmomatic [178] is a Java package that implements a number of trimming tasks, based on length and quality of each sequence, for the Illumina NGS output sequencing [56]. Bowtie 2 [98] and STAR [99] are tools for aligning sequencing reads to long reference sequences [96][97]. For example, data are firstly analyzed by FASTQC that gives objective parameters based on both positional or quality level criterion for removing low-quality reads. These parameters are then set in Trimmomatic that execute the quality filtering of each available read thus reducing the error probability of the sequences output and thus increasing the accuracy of the subsequent alignment step that maps quality-filtered data to a reference genome. Although this task could be accomplished by BLAST, it is not specialized for handling a very large amount of data generated by NGS machines. STAR and Bowtie 2 are two examples of specialized tools for such purpose. A comprehensive survey of modern aligners and the relevant comparison can be found in [96].

Among genomics tools, an important role is played by those devoted to efficiently query large databases.

GenomeTools [179] is a software library designed for developing bioinformatics software intended to create, process or convert annotation graphs. It offers a unified graph-based representation, providing the developer with an intuitive access to genomic features and tools for their manipulation. In order to process large annotation datasets with low memory overhead, GenomeTools is based on a pull-based approach for sequential processing of annotations.

In [184], the authors propose to organize genomic processing software into layers like the networking protocol stack. These layers are instrument layer, compression layer, evidence layer, inference layer, and variation layer. This layered organization can insulate genomic applications from sequencing technology. The Genome Query Language (GQL) proposed in [184] is a specific interface between the evidence and the inference layer. The evidence layer is thought to be implemented by a large cloud computing deployment to provide, for instance, a query interface that can return the subset of reads supporting specific variations. Instead, the inference layer, which may be computationally intensive, typically works on smaller amounts of data (filtered by the evidence layer), thus it can run either on the cloud or on client workstations. A genome repository accessible by GQL offers the ability to reuse genomic data across studies, the ability to logically assemble case-control cohorts, and the ability to rapidly change queries without ad-hoc programming.

The Genomic and Proteomic Knowledge Base (GPKB) [181] is a database able to integrate the most relevant sources in terms of genomic and proteomic semantic annotations, which are scattered in many distributed and heterogeneous data sources, hampering the researchers' ability of doing global queries and performing global evaluations. GPKB uses a flexible, modular, and multilevel global data schema based on abstraction and generalization of integrated data features, and a set of automatic procedures for easing data integration and maintenance, also when the integrated data sources evolve in data content, structure, and number.

The GenoMetric Query Language (GMQL) [182] is a new query language for genomic data management that operates on heterogeneous genomic datasets. The GMQL can be executed in a parallel fashion, and two different implementations with two emerging frameworks for data management on the cloud, namely Flink and Spark, are available with similar performance. The GMQL poses as a novel solution to integrate the information extracted from sequencing operations, providing holistic solutions to the needs of biologists and clinicians.

Finally, we briefly review the introduction of machine learning techniques to address genomics analysis [183]. Machine learning can help to model the relationship between DNA and the quantities of key molecules in the cell (cell variables), which may be associated with disease risks. Since nowadays high-throughput measurement of many cell variables are possible (e.g. gene expression), these can all be used as training data for predictive models. In more detail, machine learning can be used to infer models that are capable of generalizing to new genetic contexts. This notion of generalization is a crucial aspect of the models that need to be inferred. An important aspect of model development is validation using DNA sequences and cells states different of those used for training. Since it is not feasible for a model to be accurate for any input, the validation procedure should characterize the inputs for which it can be considered reliable. If a model is enough general, it could indicate a disease without needing experimental measurements, by simply analyzing mutations that change cell variables.

A very important application field of machine learning techniques in genomics is genome-editing modeling, which can be used to understand the long-term effects of genome changes. An issue related to these approaches is the management of mutations causing a large change in cell variables without implying any disease, or false negatives, which occur for mutations that act through cell variables that are not being modeled. Both errors indicate inaccuracies in the developed models. Finally, neural networks seem to provide quite inaccurate predictions when fed with adversarial inputs, that is inputs explicitly designed to "fool" a model so as it makes wrong predictions [183]. Although adversarial inputs may not occur naturally, they are important to predict the effect of therapies that make small changes to the genome, for example using genome editing technologies, since the resulting genome sequences may be unnatural. Thus, the question of testing for adversarial input arises. To address it, the validation procedure of computational models may be quite complex, requiring synthesizing adversarial genomic variants and compare predictions to real experiments.

## C. A Taxonomy for Genomic Processing Platforms

Now we present two taxonomies for genomic processing pipelines, which classify them from two different viewpoints. The first taxonomy accounts for the computing environment where a genomic pipeline is executed, whereas the second taxonomy deals with the programming approach used to implement the pipeline components. Clearly, the two taxonomies are strictly related each other.

### 1) Alternatives for genomics pipelines implementation

The taxonomy shown in Fig. 2 considers the most common alternatives used to implement genomic processing pipelines, and the relevant mapping on computing infrastructures. According to [187], the strategies used to organize and manage genomic processing pipelines can be classified into three main alternatives:

- those based on scripts,
- those relying on makefiles,
- those organized into workflows.

Scripts, written in Unix shell or other scripting languages as Ruby or Perl, are the most basic forms of pipeline implementation. They are compact and tailored for running commands in a specific order. Scripting allows using variables and conditional logic to build flexible pipelines. However, in terms of "robustness", scripts can be quite weak. In fact, due to the dependency between upstream and downstream files, upon a change upstream, all the relevant tasks have to be updated manually. Also, upon a failure during the execution of a pipeline, usually it is not possible to restart the execution from where it was interrupted, and it is necessary to re-execute the pipeline from scratch.

*Makefiles* can be used to manage file transformations in scientific computing pipelines by means of the *Make* tool [187]. Make introduced the concept of 'implicit wildcard rules', which define available file transformations based on file suffixes. By these rules, Make is able to generate a dependency tree, which allows inferring the steps required to build any target for which a rule chain exists. However, makefiles are often not flexible enough, since they do not support multi-threaded or multi-process jobs for exploiting underlying parallelization capabilities, and they do not provide any means to describe a recursive flow.

To sum up, makefiles are good for simple pipelines applied to basic use cases, but become unsuitable when the pipelines get more complex, with multiple steps and branches.

Finally, in recent years a number of modern pipeline frameworks have been developed to address the Make's limitations in syntax, monitoring, and parallel processing. In addition, these frameworks, commonly referred to as workflows, provide bioinformatics with new features, such as visualization, version tracking, and summary reports. A detailed discussion of the different flavors of workflows are discussed in [186]. According to [186], we can classify workflows according to

- the type of syntax (*implicit*, like Make, or *explicit*, more similar to scripts);
- the programming paradigm (based on *convention*, which uses inline scripting codes, *configuration*, which requires configuration files, e.g. in XML, or *classes*, which relies mainly on code libraries and not on executable files);
- the interface (*command line* or *graphical*).

Combining the different viewpoints of [186] and [187], we adopted the interface criteria to differentiate pipelines, grouping together modern workflows implemented by command line programs, e.g. written in Python, with scripts and makefiles. In fact, not only it is the most immediate way to discriminate how pipelines are built and managed, but they also have many commonalities on the hardware architecture where they can be executed. Fig. 2 shows the proposed taxonomy. In more detail, when we consider the architecture used to run these pipelines, command line approaches can be run into a single server, in a virtual machine (VM), in a platform-as-a-service (PaaS) cloud environment, or in a distributed environment, implemented by means of a local cluster, a grid, or a HPC system. In turn, the distributed environment can be implemented by physical servers, or can be deployed by using VMs on top of IaaS public or private clouds.

As for the two first categories, we can mention Bio-Linux [188][189]. It is a free platform that can be installed on any type of computing device, or run as a VM according to the IaaS paradigm. Currently, it implements more than 250 bioinformatics packages, with about 50 graphical applications and several hundred command line tools. In [5], the authors presented a web interface for using the genomic software in the Bio-Linux VM, according to the SaaS cloud paradigm, thus on top of an IaaS deployment. Other examples of public IaaS services, used for running VMs with genomic processing software, exist. They include Amazon web services (AWS) elastic cloud computing (EC2) [16][106] and Microsoft Azure [108] cloud services.

As for the command line PaaS environment for developing genomic pipelines, a noticeable example is given by the Google Genomics platform [190]. It provides an ample set of libraries [110], which can be used to efficiently implement pipelines by means of scripting languages [109], also exploiting the presence of genomic reference files [111], as well as a copy of the 1000 genomes datasets [210] in the Google storage. In turn, these pipelines run in scalable clusters of VMs, depending on the user requested resources.

Finally, we consider command line pipelines (mainly scripts or modern workflows), which can run in distributed environments, and are designed essentially to automate the pipeline development and management process. The paper [191] provides an interesting tutorial about the execution of genomic pipelines in distributed environments, encompassing clusters, grids, and HPC infrastructures. When dealing with command line tools, a well-known workflow management system is Pegasus [193], which can run in multiple distributed environments, from clusters to grids to HPCs systems, either deployed on physical machines or virtualized in a cloud environment. It is a configuration-based framework, and it requires a configuration XML file that describes individual job instances and their dependencies. An example of a genomics pipeline implemented with the Pegasus workflow and running in a grid environment is the OSG-GEM [195], which leverages the computing resources of the NSF/DOE Open Science Grid (OSG). PGen [194] is an example of a pipeline implemented by using Pegasus, but running in a virtualized HPC environment, the Extreme Science and Engineering Discovery Environment (XSEDE), which is the federation of supercomputers supported by the NSF. Another example of command line workflow tool, implemented in Python and able to run in virtualized clusters, is Kronos [192], which can leverage the virtualization capabilities of both Docker containers or AWS machine images. It avoids writing the code of workflows, as it requires just to compile a text configuration file into executable Python applications. A portable system, used for expressing and running a data-intensive workflow without requiring changes to the application or workflow description, is Makeflow [200]. It is inspired to the Unix Make, thus easy to use by users familiar with makefiles, and improves the Make parallel execution support, since it is able to run smoothly on single machines, local clusters (e.g. managed with Condor), and grid systems, either physical or virtualized in cloud environments.

Finally, for what concerns graphical workflows (or workbenches [186]), we have identified two main categories: those based on PaaS, which offer not only graphical tools to assemble pipelines but also development tools, and those based on distributed virtual environments, where it is essentially possible to assemble pipeline by just graphically combining available components, without any specific support for component development.

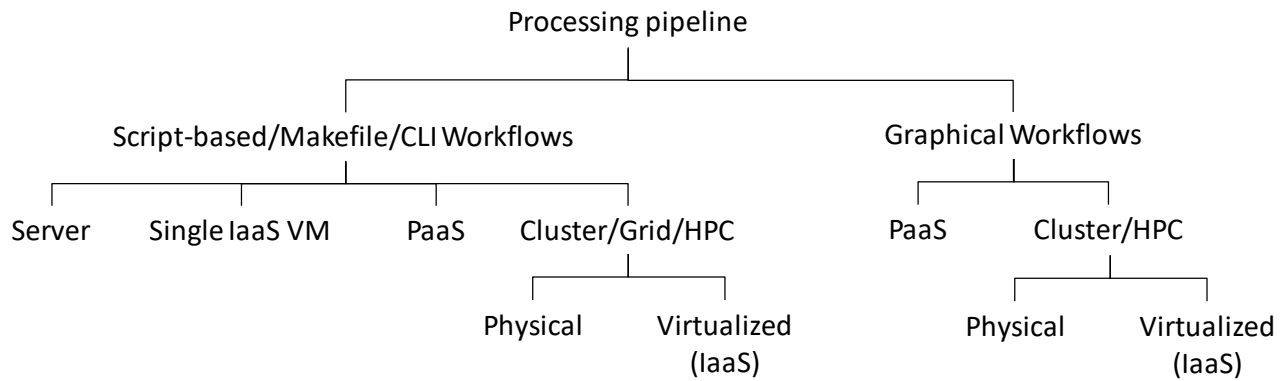

Fig. 2. Taxonomy of computing infrastructures for genomic processing.

In the first category, we can find DNAnexus [196], which operates on top of AWS virtual machines. As for the latter one, we can mention Galaxy [197], which can run both in physical and virtualized clusters, and provides a Web-based interface able to mask the underlying (command line) tools [186].

Another alternative is the Graphical Pipeline for Computational Genomics (GPCG) [199], which is a collection of "ready to use" workflows covering a broad spectrum of DNA-Seq data analysis steps. In GPCC, the developers have converted a number of command line processes into module definitions, and then connected them to form workflows.

As a general comment, for the general adoption of more advanced workflows, a significant issue is the lack of standardized data flow. Up to now, no widely accepted approaches exist in the bioinformatics community [187]. Clearly, research workflows created to run ad hoc analysis cannot be standardized. In fact, in these situations steps and parameters of pipelines are often modified, as the understanding of the issues being investigated progresses. For what concerns more common workflows, also in this case many scientists prefer to develop tools in-house, so as to adapt them to their specific scenarios. In addition, adding new modules/functions in workflow tools may require an in-depth knowledge of the platform, especially when dealing with graphical workbenches. Finally, workflow tools from other domains, such as astronomy, are typically not used, not only due to lack of communication between different research fields, but also due to domain specific needs.

### 2) Programming approaches for genomics pipelines components

The second classification, which is related to the first one, is relevant to the software technologies that can be used to implement pipelines or, more often, their specific components, such as alignment or SNP detector tools. Fig. 3 shows the available alternatives to implement these components. We identified four main alternatives, which are classic programming using threads on CPU, graphical processing unit (GPU) based-programming, exploitation of the processing capabilities of dedicated co-processors, or using the MapReduce paradigm.

As matter of fact, there are still many popular genomics analysis tools which have been designed as mono-thread software programs. This is due to two main reasons. The first one is that the initial version of the program could have been implemented in the most straightforward way, that is single thread, and immediately released without any further optimization. It happens when the main focus of the activity is the algorithm processing genomic data, and not on its optimized implementation. The second is that, often, the specific processing that these modules execute on the genomic data is intrinsically sequential, which makes parallelization of its execution not easy. Some notable examples of this type of software components, extensively used in genomic pipelines (see also the examples in the next section V.D and the general description in section V.B), are CNVnator [93], a tool for CNV discovery and genotyping from depth-of-coverage by mapped reads, and htseq-count, a tool developed with HTSeq [105], that pre-processes RNA-Seq data for differential expression analysis by counting the overlaps of reads with genes. In conclusion, what often happens today is that these genomic packages cannot exploit the computing capacity of even single servers. In addition, due to the need of minimizing the consumed power and energy, the design of processors has evolved towards multi-core architectures.

Thus, if a genomic software package cannot be easily parallelized, this issue is found even if it is executed in a single server. Thus, if these software packages become popular, it could be very difficult to replace them with a more performing version, able to exploit the intrinsic parallelism of modern multi-core architectures.

Nevertheless, since in some cases these tools are the main performance limiting components during pipeline execution, when the volume of genomic data increases, an effort is needed to improve their implementation toward parallelization. This is the case of HTSeq, for which a more efficient multi-thread implementation named VERSE [203] has been released. VERSE can decrease the computation time by more than 30 times when executed on the same multi-core machine.

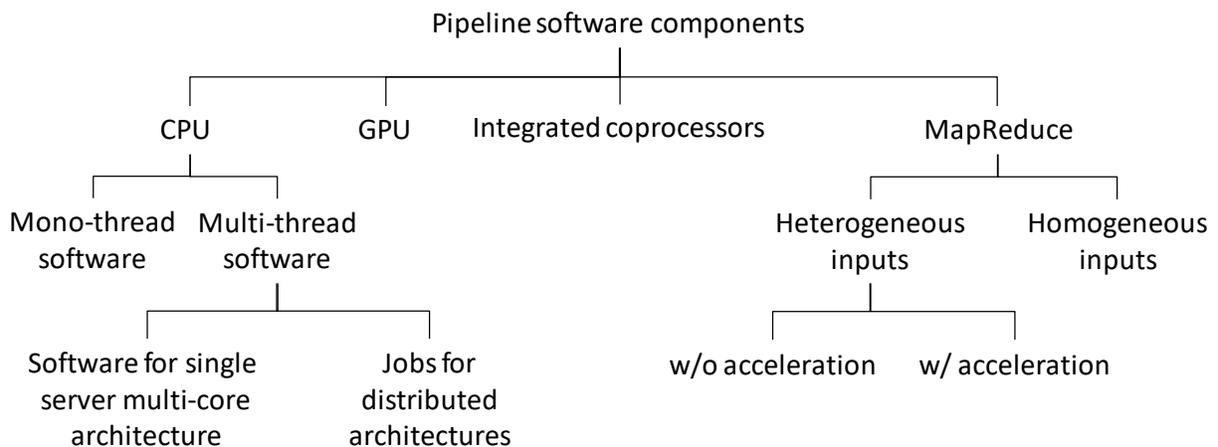

Fig. 3. Taxonomy of implementation approaches for genomic pipeline components.

A further effort toward a higher level of parallelism is the porting of the execution environment of each software component in a distributed environment. This can be done in two different ways [204]. The first one is the classic redesign of bioinformatics tools (e.g., BLAST) to their parallel versions using message passing interface (MPI) or similar approaches. For instance, the previously mentioned mpiBLAST [34] can achieve super-linear speed-up with the size of database. A second strategy can rely on the same tools used to develop entire genomics pipelines, that is workflows.

An interesting example is given by Makeflow [200], which is able not only to execute pipelines in distributed environments, but even to split the execution of a single pipeline component by means of a technique known as job expansion [202]. When job expansion is applied to a pipeline workflow, each node in the logical workflow is expanded into another, large workflow with potentially hundreds to thousands of tasks that can be executed in parallel, thus enabling high concurrency and scalability. This process is completely transparent to the user. With reference to BLAST, its set of query sequences can be split into multiple tasks, each one able to run its assigned query sequences against a duplicated copy of the reference database [200].

Other possibilities for handling the increasing volume of genomic data are represented by programming techniques for specific hardware architectures, and more specifically (i) dedicated coprocessors with many integrated cores (MIC), such as the Intel Phi [201], and (ii) graphical processing units (GPU) [206]. An interesting performance comparison of a bioinformatics application over CPU, GPU and MIC, can be found in [205]. It results that the approaches leveraging both MIC and GPU capabilities can significantly outperform classic CPU-based processing. In addition, GPUs result significantly more efficient than MICs when data access patterns are mainly based on random access, corresponding to operations that access data irregularly. Actually, some initiatives aim to port genomics tools in GPU environments. The usage of GPU is particularly appealing when it is possible to leverage high levels of parallelism. In the framework of genomics analysis, it happens in alignment tools. In fact, since NGS high throughput machines usually produce a huge number of subsequences (billions of so-called 'short reads') of the target genome, they need to be realigned with a reference sequence. This operation can be efficiently carried out by using the many core processors of a GPU. For this purpose, one the most frequently used GPU architecture is the NVIDIA compute unified device architecture (CUDA). For example, CUSHAW [29] and BarraCUDA [207] use CUDA. The work [206] presents a quite complete list of these software tools.

An additional option to speed up the alignment phase uses a different type of specialized processors, the Field-Programmable Gate Arrays (FPGAs). In particular, the results presented in [208] proved that FPGA-based alignment is able to outperform CPU-based processing (Bowtie2 [98]) by 28 times, and a GPU-based implementation by 9 times. While this approach seems really promising, having solved the accuracy issues present in previous FPGA-based proposals, it comes with still unsolved issues, which are mainly related to the I/O operations with the FPGA board. In fact, in the evaluation presented in [208], the authors state that they have purposely omitted not only the FPGA reconfiguration time, which is quite negligible (few seconds), but also the time needed to pre-load the reference file and the disk I/O time, to avoid negatively biasing the processing capabilities of their approach. Thus, a complete performance comparison with other acceleration techniques is still missing.

The need of managing large volume of data makes the Google MapReduce computing paradigm [134] a possible candidate baseline for distributed processing of genomic data. It leverages on the computational power achievable by parallel processing by using many computing devices simultaneously. Large computational tasks are repeatedly broken down into small portions that are distributed to individual computers across the network. When individual jobs are accomplished (map phase), individual results are grouped and aggregated (reduce phase). Unfortunately, as mentioned above, the mostly used genomic processing tools cannot be easily parallelized, since not only they have essentially been designed to be executed in very powerful stand-alone servers, but they are implemented also as mono-thread software, thus unable to exploit the computing capacity of even single servers. This makes it even more difficult if the processing is distributed over multiple servers. However, this trend is evolving, with more and more genomic applications being made complaint with Apache Hadoop [55][120][122], the open-source implementation of the MapReduce framework, or other proprietary MapReduce frameworks, as discussed in [9], or other more modern and efficient Big Data processing engines, such as Flink and Spark [182]. Table 1 in [9] reports a list of genomics software packages already implemented and available in the MapReduce processing framework. However, not all genomic programs have been or can be ported. Thus, in order to implement a specific pipeline, it can happen that part of the available software is still implemented with classic programming paradigms, with limited parallel processing support. In addition, as also discussed in [121], "Although there are many successful applications of Hadoop, including processing NGS data, not all programs fit this model and learning the Hadoop framework can be challenging." Finally, the use of Hadoop is appropriate for processing extremely large volumes of data, when it is really able to take advantage of its parallel processing power. Instead, for executing individual genomic services coming on request (e.g., making use of a few genomics read files, the size of which is significant but not massive), it could be not the most cost-effective solution. In Fig. 3, we identified two cases: inputs from the same source (homogeneous), and inputs from different sources (heterogeneous). Thus, small homogeneous requests could not fit well the MapReduce paradigm, and need to be aggregated into heterogeneous inputs. In this regard, CloudBurst has proved that MapReduce is also an excellent paradigm to speed up the short-read mapping process by exploiting its intrinsic parallelism [167]. However, CloudBurst does not begin the service until millions of short reads have been loaded into the system. This process can be extremely time-consuming, and any update to reads would require a complete re-execution of MapReduce. To help mitigating this issue, in [166] the authors propose a novel FPGA-based acceleration solution with MapReduce framework on multiple hardware accelerators, so as to speed up the processing also in the case of heterogeneous inputs.

Finally, Hadoop allows also managing data storage over computer clusters through the Hadoop Distributed File System (HDFS). This feature opens up new possibilities for managing distributed computing, exploited in recent projects. For instance, in [209] the authors demonstrate the scalability of a sequence alignment pipeline carried out by using Apache Flink and HDFS, both running on the distributed Apache YARN platform. The novelty is that their pipeline is able to directly process with Flink the raw data produced by the NGS sequencers. Subsequently, a distributed aligner is used to perform read mapping on the YARN platform. Results in [209] show that the pipeline linearly scales with the number of computing nodes. In addition, this approach natively brings in benefits from the robustness to failures provided by the YARN platform, as well as the scalability of HDFS. Moreover, it opens new possibility for using other software packages compliant with YARN, paving the way towards distributed in-memory file system technologies such as Apache Arrow, able to remove the need of writing intermediate data to disk.

### D. Experimental assessment of genome analysis pipelines

After having identified some widely used genomic software tools, and classified the commonly used computing approaches for implementing genomic processing pipelines, this section shows two typical pipeline examples, implemented according to the paradigm *script-based pipeline* deployed in a *single IaaS VM* with multi-core computing capabilities running both *mono-* and *multi-thread software packages*. We carried out some experiments that allowed us to get an insight about processing time and computing resource consumption. Although this analysis cannot be inclusive of all uses of genomic data sets, it can provide some basic understanding of the performance behavior of genomic tools in terms of processing time and resource requirements for the specific paradigm under evaluation, thus providing hints for a more general workload characterization. Further implemented pipelines are illustrated in [115].

The input files of the genomic pipelines shown in this paper are:
- FASTQ files [103], containing sequencer output raw data;
- FASTA files [104], containing typically quality-filtered sequences (genome, exome, or transcriptome);
- annotations files [20], containing lists of gene sequences.

These software pipelines are managed by a ruby script running in a VM executed by the KVM hypervisor in the Linux operating system (OS), in a private cloud managed by means of the OpenStack cloud management software [137].

The first pipeline, shown in Fig. 4, implements the CNV analysis. It includes some of the software tools mentioned in section V.B. FASTQ files have been taken from the *1000*

*Genomes* repository [100]. The first element of the pipeline is the Trimmomatic package, which performs quality trimming on genomic reads. Essentially, each base of the sequence read of the input FASTQ file is checked and filtered based on its Phred quality score, an integer value that measures the quality of the single nucleotide identification and that is related logarithmically to the error probability identification (i.e. *Phred* quality score is the ratio, expressed in dB and with changed sign, between the error probability on the single nucleotide identification and the unitary probability level). Phred quality score values are encoded with ASCII character based on different algorithm (Phred+33, Phred+64, Solexa+64) depending on the sequencing platform and represent the reliability of the sequencing procedure. If the overall quality of a line does not exceed a desired threshold it is discarded or trimmed. The "filtered" file is passed to the Bowtie 2 [98], which aligns it with the latest human genome reference model, the *human genome* 19 (hg19), available on-line [101][102]. The actual CVN analysis is performed by CNVnator [93], the output of which is used to produce custom reports.

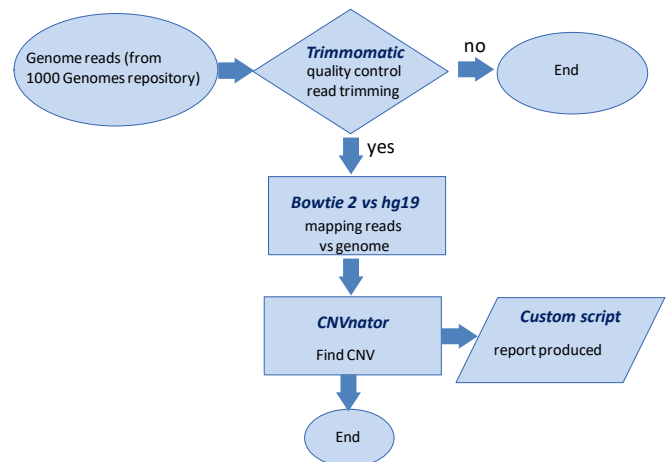

Fig. 4. CNV pipeline.

The second example is the DE processing pipeline illustrated in Fig. 5. The initial step for quality control is like that of the CNV pipeline. The alignment of the trimmed FASTQ file with the hg19 file is done by using STAR [96]. The actual differential expression analysis is executed by using the gene expression counting functions of HTSeq and the subsequent comparison implemented by a custom script written in R. A further log file, including information about all processing steps, is produced.

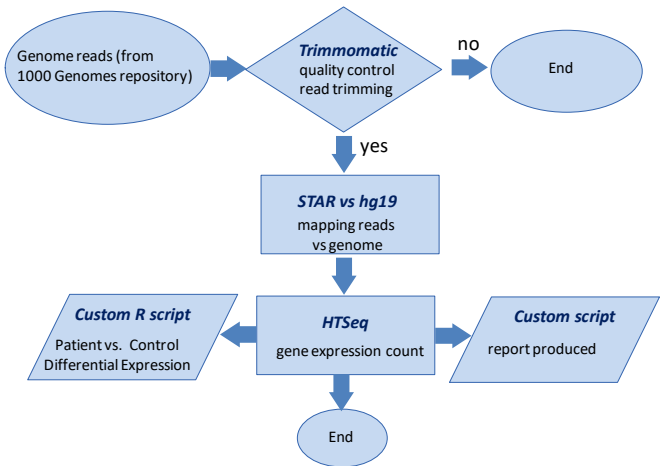

Fig. 5. DE pipeline.

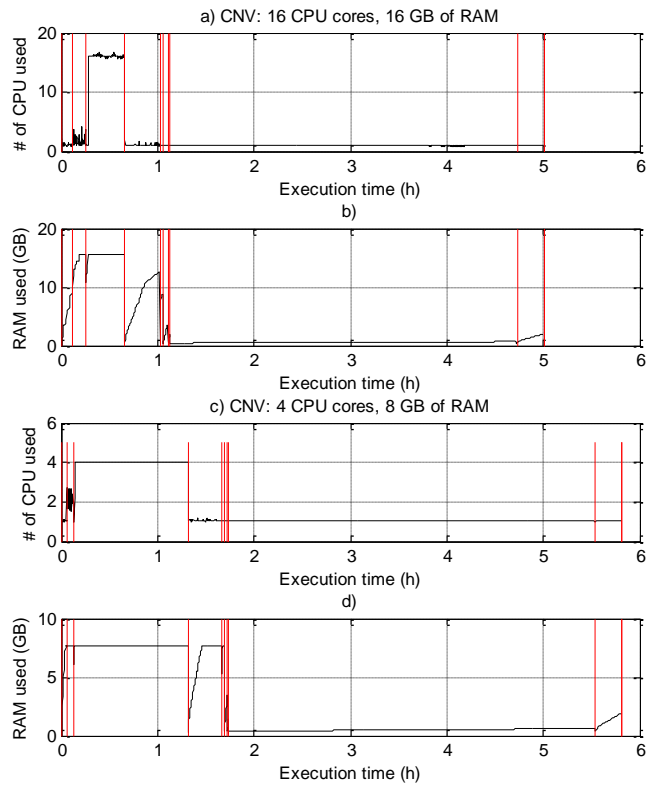

Fig. 6. CPU and memory requirements for a pipeline implementing a CNV analysis, with two configurations: a) 16 CPU core, 16 GB of RAM, and b) 4 CPU cores and 8 GB of RAM. The input file, taken from the public repository of the 1000 Genomes Project [210], is about 8 GB.

Table 2 reports some different configurations for the two pipelines, and the relevant minimum requirements in terms of amount of RAM, number of CPU cores, and virtual disk size to correctly execute them. The input size and the auxiliary files size (e.g. the genome indexes) reported in Table 2 represents the total size of the files to be loaded into the VM in order to the execute the relevant genomic pipeline. Even a case of DE implemented by using the Bowtie 2 aligner is reported for completeness, although not analyzed in-depth in the analysis shown in what follows.

Before illustrating some results of the experimental analysis of the execution time in different case studies, it is necessary to have an idea about how the implemented pipelines make use of the computing resources in the different steps of their execution.

Fig. 6.a shows an example relevant to the CNV pipeline, with a VM with 16 CPU cores and 16 GB of RAM. It shows how the CNV makes use of the available computing and memory resources, for a relatively small input size (patient file) of about 8 GB.

TABLE 2: CONFIGURATION AND MINIMUM RESOURCE REQUESTS OF VMS IMPLEMENTING THE GENOMIC PIPELINES.

| Pipeline | Configuration | Hypervisor | VM image size | Min RAM size | min # CPU cores | VM storage size [1] | Auxiliary files size |
|---|---|---|---|---|---|---|---|
| CNV | BOWTIE aligner[2] | KVM | 3.1GB | 8 GB | 1 | 50 GB | 3.5 GB |
| CNV | BOWTIE aligner[3] | KVM | 3.1 GB | 4 GB | 1 | 50 GB | 3.5 GB |
| DE | BOWTIE aligner | KVM | 3.1 GB | 4 GB | 1 | 80 GB | 3.5 GB |
| DE | STAR aligner | KVM | 3.1 GB | 32 GB | 1 | 100 GB | 26 GB |

TABLE 3: INPUT AND REFENCE FILES USED IN CNV AND DE EXPERIMENTS (GENOMIC READS FROM 1000GENOMES [100]).

| Parameters | Experiment 1 (CNV pipeline) | Experiment 2 (DE pipeline) | Experiment 3 (DE pipeline) | Experiment 4 (DE pipeline) |
|---|---|---|---|---|
| Patient | HG00097 | HG00097 | HG00259 | HG00334 |
| Input files (GB) | SRR741385 1 (7.1) SRR741385 2 (7.0) SRR741384 1 (6.1) SRR741384 2 (6.1) | ERR188231 1 (3.15) ERR188231 2 (3.18) | ERR188378 1 (3.70) ERR188378 2 (3.73) | ERR188127 1 (3.70) ERR188127 2 (3.73) |

---

[1] The requirement in terms of storage size allocated to the VM executing the processing pipeline (in GB) has been estimated by using as inputs two compressed files of 1.2 GB each. It takes into account all intermediate outputs of the pipeline, and leaves enough spare disk space to avoid problems in the operating system.

[2] Computing is performed on the whole human genome.

[3] Computing is performed chromosome by chromosome.

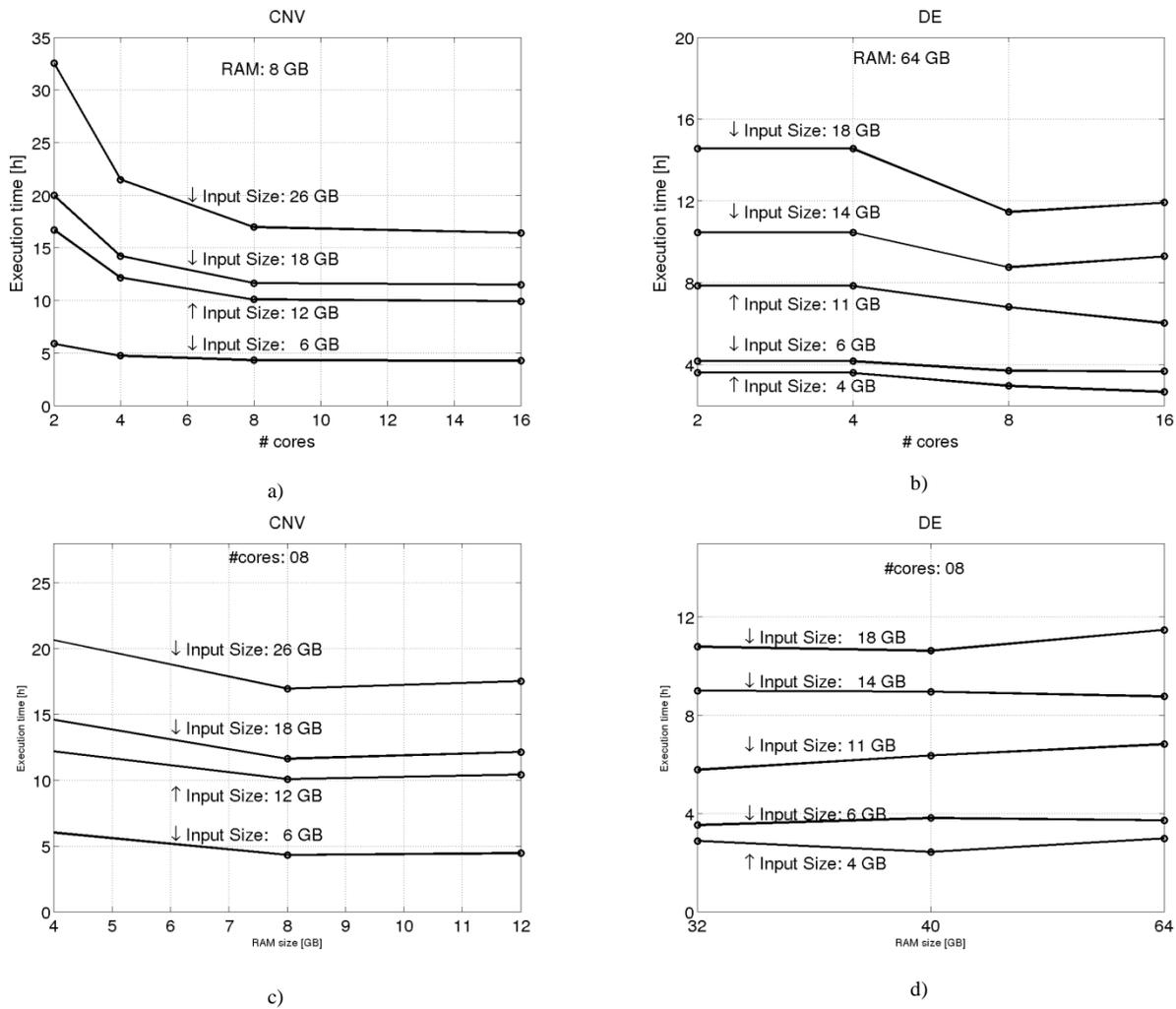

Fig. 7. Experimental results for CNV and DE pipelines: execution time versus computing resources (number of CPU cores for subfigures a and b, amount of RAM for subfigures c and d), with approximated input size as curve parameter. The amount of memory is 8Gb for CNV and 64 GB for DE, respectively, in subfigures a and b. The number of CPU cores is 8 for both CNV and DE in subfigures c and d. Input data are publicly available in [100]. See also Table 3.

Vertical red bars delimitate different sub-phases within the processing. It is evident that the pipeline is quite inefficient in the usage of computing resources. Just in the initial phases the multi-core architecture is exploited, as most of software packages included in the pipeline just make use of a single core. Similar considerations can be done also for what concerns the amount of RAM used. It is arguable that the reason of this behavior is that, initially, the bioinformatics programmers had not the need of using multi-core machines. Nevertheless, given the increasing diffusion of multi-core architectures, including both CPUs and GPUs, this issue is now significant and needs a specific solution, especially because we have used one of the most popular genomic processing software, CNVnator [93].

In Fig. 6.b we can appreciate a different behavior when the computing resources of the VM are reduced to 4 CPU cores (4 times less) and 8 GB of RAM (halved).

As we can see, just the first phase (alignment), which is the most demanding from a computational viewpoint, is able to exploit the underlying multi-core computing architectures, and thus it presents a longer processing time (about 4 times longer), which results inversely proportional to the number of CPU cores.

The remainder of the elaboration, being carried out with single-thread software, does not exhibit any slowdown due to the reduced amount of computational resources.

In addition, we have observed a very similar behavior even in the DE processing. Hence, a first general comment is that the number of cores and the RAM size allocated to VMs should be dynamically updated during pipeline execution, in order to avoid waste of resources.

Processing time is one of the most important performance figure for genomic processing. Thus, we have investigated the relationship between computing resources and processing time.

Fig. 7 shows some sample results of an extensive experimental campaign using the CVN and DE pipelines. We report the name of the genomic read files downloaded from [100], the combination of which allows obtaining the input size for each curve. The execution time is shown as a function of both the number of CPU cores (Fig. 7.a and Fig. 7.b) and the amount of RAM allocated to the processing VM (Fig. 7.c and Fig. 7.d). As already discussed in the analysis of Fig. 6, increasing the number of CPU cores affects only the initial processing phases of the pipelines. Thus, the consequent performance improvement in terms of execution time is limited. In more detail, it has a significant impact on service time only when the allocated CPU cores increase from 2 to 4. Once the duration of alignment phase is reduced enough, deploying additional computing resources is quite useless.

What emerges from Fig. 7.c and Fig. 7.d is that both CNV and DE pipelines are quite unaffected by the allocation of an amount of RAM larger than the minimum requirements. This rises further suspects that the software design is inefficient also with respect to the memory management.

The insight from these results is that for introducing different service classes, distinguished by the service time, the most effective approach should consider the input sizes of genomes. In fact, the execution time scales almost linearly with the input size, whilst it is quite unaffected by over-allocation of RAM and CPU cores.

This conclusion could be reconsidered when most of bioinformatic operators will make use of the highly parallelized genomic programs currently under implementation (e.g. [105]). In more practical terms, it means that, when serious medical issues are handled, the patient's genome has to be processed individually, since the processing time of aggregated multiple genomic reads, as it is typically done in genomic computing, cannot be significantly reduced by over-allocating computing resources. In case of very large volume of genomic data to process, a viable alternative to significantly reduce the processing time is the usage of Hadoop [9] or other big data engines [182], in case the whole desired pipeline is implemented according to the MapReduce framework, as already discussed in section V.C.

To sum up, for the case of a single VM, we now provide a generic workload characterization based on quantitative data shown in Fig. 6 and Fig. 7. We identify the main processing phases within the VM lifecycle, and the relevant resource consumption associated with them, assuming to run the VM in a large IaaS deployment. Fig. 8 sketches the allocated resources in terms of computing resources (CPU and memory, top subfigure) and networking resources (bandwidth, bottom subfigure) of a generic service request for an IaaS paradigm. The time duration of each phase are qualitative and not in scale. However, they provide a qualitative indication about the relevant weight of the phases.

In the first phase, we assume that the VM disk image containing the genomic processing is not present in the local VM image repository (worst case). Thus, it is necessary to download it from an external repository. Thus, a given amount of computing resources is reserved, although during the initial phase just the VM image is downloaded (yellow area). Assuming that the size of the compressed VM image, with all installed software, is a few GBs, and the throughput for downloading it from a remote server is in the order of a few hundreds of Mb/s, the time needed to complete this phase is in the order of few minutes. Instead, the time needed to boot the VM is usually in the order of few seconds. An alternative approach could be to use a clean cloud OS image and install the software at the first boot by using the *cloud-init* approach [236]. In this way, since the VM image will be available in the local image repository with very high probability, the remote download will be avoided. However, at the first boot of the VM, the scripts launched by cloud-init could require to download and configure a significant amount of software packages, thus significantly increasing its duration. In general, it is not easy to predict which is the fastest solution. In addition, by using cloud-init scripts, this process should be repeated at each VM instance creation.

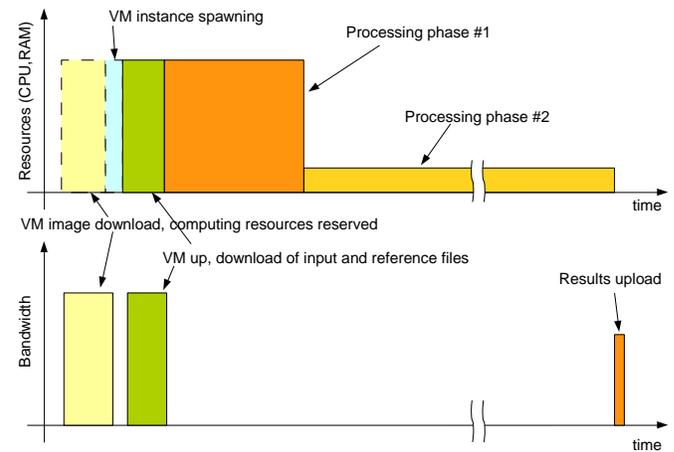

Fig. 8. Workload in terms of CPU, memory, and bandwidth for a typical processing request. Different phases (download and/or processing) are highlighted. Timescales are indicative.

After the VM is started, most of the computing resources (green area) are reserved, but may be still unused. In fact, during this phase, the VM itself needs to retrieve input files (e.g. genomes) and eventually reference files, if not already included in the VM image disk. Also in this case, depending on the size of the files to download (a few to several GBs) and the download speed, this phase could last from few to several minutes.

At the end of this phase, the service may not need to access the network anymore[4]. Typically, the initial processing phase is resource intensive (e.g. alignment of the input genome with respect to a known sequence, as discussed in section V.C). This phase is highlighted in orange and marked as "Processing phase #1" in the figure. As it appears in Fig. 6, the length of this phase is strongly dependent not only on the type of processing, but also on the amount of computing resources available to the VM. A typical duration is in the order of tens of minutes, thus definitely larger than the previous ones. It is followed by another phase, "Processing phase #2", typically characterized by much lower computing resources requirements, during which computing resources can be resized by using a dynamic scheduler for cloud environments [211]. The duration of this phase, depending on the processing type and the adopted computing approach, as well as the amount of data to process, can last even some hours.

---

[4] In case the used software does not include downloading auxiliary reference files but querying a remote database, this network activity can be included in the phase corresponding to the "green area".

Finally, at the end of the computation, processing results are uploaded to a specific location, accessible to the end user. However, the typical size of the output is in the order of a few MBs, thus orders of magnitude lower than the inputs.

## VI. DATA NETWORKS FOR GENOMIC COMPUTING

### A. Functional features for genomic networking solutions

Before illustrating the currently available solutions for genomic networking and discuss their potential adoption in a GaaS implementation, it is useful to contextualize their features through a suitable classification for comparing them and for guiding future research. We consider the following scheme to define the functional features of a GaaS platform:

- *Scalability*: Similarly to other Big Data contexts, the ability of network services to handle a growing amount of traffic is essential. Nevertheless, scalability of genomic networking should be regarded in two dimensions, namely the overall traffic exchanged through networks and the increasing traffic involved in each individual service instance, which typically involves multiple genome reads, reference genome files and metadata.
- *Flexibility*: Ease of adapting to different applications contexts, for either research, medical, and any other business applications. DNA processing is increasingly used in many fields, and network service should be flexible enough to allow supporting a plethora of applications efficiently.
- *Specificity*: Designed so as to suitably exploiting the intrinsic features of genomic data set and application practices.
- *Security*: Authentication, encryption, integrity of contents, and confidentiality are needed.
- *Deployability*: Intended as being easily deployable by using publicly available software components, preferably open source, with a short time from design to production, and able to providing inter-operator service deployment. This feature is pursuable through recent achievements in the networking area, such as *programmable networks* and *network function virtualization*.
- *Optimality*: Networking protocols should be driven by optimization algorithms targeted to the specific genomic services, for example for discovering and managing the available resources, such as caches, bandwidth, and computing facilities, according to the service needs.
- *Accessibility*: This feature is borderline in regard to the networking services. It refers to the application front-office that allow access to services automatically, in order to allow any user to successfully exploit all the available potentials. Cloud-based access to services is a suitable approach [5].
- *Efficiency*: Network solutions should be designed to increase as much as possible the utilization of deployed network resources, in order to maximize the benefit of invested capitals. Since different solutions, among those considered in VI.B, use different approaches, we will compare them, highlighting pros and cons.

### B. Existing networked genomic solutions

Some computational platforms, realized for supporting general services eager of storage space and processing power, are currently available [19]. The AWS EC2 service [44] allows users to rent VMs to execute their applications. VMs are characterizes through *compute units*. For what concerns storage space for handling the Big Data problem, the Amazon S3 cloud computing service [45] provides an interface for web services to store and retrieve, namely, any volume of data. As mentioned before, IaaS services such as those offered by AWS can be used both to run individual VMs, such as Cloud Biolinux [5], and to deploy more advanced bioinformatics service frameworks, such as Galaxy [197].

Although these general services constitute a valuable support for most of the current needs of genomic processing, some concerns emerge in regard to their suitability for sustaining either the computing needs of very specific massive analysis or the expected generalized use of genomic processing. For example, the Translational Genomics Research Institute (TGen) hosts a cluster of several hundreds of CPU cores to significantly decrease the processing time of the tremendous volume of data relevant to the neuroblastoma [14]. Clearly, although this brute force approach may be replicated in a small number of prestigious organizations, it cannot be generally adopted when genomic processing will be largely needed in most of countries.

Two international initiatives currently offering their genomic processing services to researchers are the Elixir organization in Europe and the Bionimbus Protected Data Cloud (PDC) in the USA. Elixir [24] is an intergovernmental organization that brings together life science resources from across Europe, such as databases, software tools, training materials, cloud storage and supercomputers. It pursues a pan-European network and storage infrastructure for biological information, supporting life science research and its translation to different fields, such as medicine, agriculture, bioindustries, and society. It allows leading life science organizations in Europe to manage and safeguard the massive amounts of data being generated every day by publicly funded research. The project Bionimbus PDC [18] is a collaboration between the Institute for Genomics and Systems Biology (IGSB) at the University of Chicago and the Open Science Data Cloud to develop open source technology for managing, analyzing, transporting, and sharing large genomics datasets in a secure and compliant fashion.

In regard to the proposed features for genomic networking, Table 4 reports a qualitative evaluation of the mentioned approaches by using the information available in the referenced literature. For what concerns the resource usage efficiency, in some situations it is not a suitable evaluation metric. For example, in the case of private clusters, such as the TGen cluster, it is not deployed in multiple sites for a public use. Thus, it is quite difficult to estimate, for instance, the usage efficiency of network resources. For what concerns general clouds, such as the ones of Amazon or Google, very few information is available about the protocols used in the network connecting datacenters. In any case, the input files can be loaded through the standard HTTP protocol, which, in some case, can be highly inefficient. However, as previously mentioned, both Amazon and Google clouds include data which are used by genomics processing. In particular, the Amazon S3 storage service includes the outcome of the 1000 genomes project [112], whereas Google provides fast access to some reference datasets [111]. Hence, both initiatives aim to reduce the traffic exchanged with the external networks, which could be considered, to a certain extent, related to the network resource usage efficiency, although users are unaware of the data storage location.

As for the Elixir project, it makes use of the Géant broadband network, but no specific protocols or architectural solutions have been introduced to achieve a high utilization efficiency of the network resources. Differently, particular solutions have been adopted by the Bionimbus project. More specifically, data exchanges in Bionimbus make use of a combined usage of an highly optimized transport protocol and an efficient application protocol, aimed to both exploiting available network resources and to save traffic. More specifically, data exchanges between datacenters in Bionimbus make use of a highly optimized transport protocol, UDT, an UDP-based data transfer protocol for high speed geographical networks [128]. At application layer, Bionimbus uses application named UDR, which is able to integrate UDT with the Unix rsync program, which transfers just the differences between two datasets across network links, thus saving network bandwidth. Recently the Bionimbus PDC has been integrated with the BioSDX facility [213], which is a programmable network infrastructure able to dynamically interconnect data sources and computing nodes to support complex workflows. BioSDX is based on the concept of software-defined networking (SDN) and software-defined network exchange (SDX), which enables SDN technologies to interconnect facilities in different network domains. BioSDX provides the Bionimbus PDC with a high degree of deployability.

TABLE 4: QUALITATIVE EVALUATION OF THE MOST POPULAR GENOMIC NETWORKING APPROACHES.

| Feature | General Clouds | Bionimbus | Elixir | Private clusters |
|---|---|---|---|---|
| Scalability | x | x | x | |
| Flexibility | x | | x | |
| Specificity | | x | x | x |
| Security | x | x | x | x |
| Deployability | x | x | N/A | |
| Optimality | | | | x |
| Accessibility | x | x | x | x |
| Efficiency | x | x | N/A | N/A |

In addition to these service platforms, there are also smaller research initiatives that have built prototypes, such as the ARES project (Advanced Networking for EU genomic research) [177], funded in the framework of the first Géant open call. However, the goal of projects like ARES is mainly to show the feasibility of advanced approaches for networked genomic processing, rather than building a platform open to other researchers and able to offer public processing services.

## C. Taxonomies for genomic networking solutions

A fundamental feature of network services for GaaS is the control of service delivery time. This may be a pressing requirement when dealing with specific service categories, such as medical services. In fact, the management of some serious medical situations may require the reception of reliable processing results by a specific time. This means that even the network services, which could have a significant impact on the overall service time, should be suitably managed. For instance, generic clouds cannot provide sufficient guarantees, due to the fact that the VM location may have a significant impact on service time, due to the time needed to upload patience genomes and other auxiliary files, if not already stored in the VM. Although some cloud providers offer users some guarantees about the geographical proximity of their VMs, this may be not sufficient for providing any optimality or service delivery guarantees. For instance, with AWS EC2 [44], users have a certain degree of control over the VM geographical location, in order to obtain some positive effects on latency. However, proximity is beneficial mainly for the user interaction with the instantiated VMs. In the perspective of genomic Big Data, bioinformatics services could require the transfer of massive data volumes and incur serious data management issues as it happens for generic Big Data transfer in clouds [107]. Thus, the optimal site for transferring such an amount of data in order to minimize the overall service time could be far from the user, and specific network orchestration solutions are needed to identify and to handle it. A network orchestrator can be defined as a control system for the provision, management, and optimization of network services [237]. It handles network service requests and, based on the available resources and the topological properties of the underlying network, it defines and executes a deployment plan that fulfills the functional and connectivity requirements of each service. In parallel, it also monitors the performance of all services to dynamically adjust the network configuration to continuously provide performance guarantees and cost objectives. The concept of service orchestrator is becoming more and more popular under the context of the forthcoming 5G services. Under this viewpoint, GaaS can be seen as a service implemented with a dedicated network slice with specific network service functions and requirements [238][239].

In addition to network delivery time, another important metric associated with GaaS, and Big Data in general, is the volume of traffic generated by networked Big Data services. This is important especially for service requests that do not pose a specific constraint on delivery deadline, such as massive processing requests for research purposes. With the expected tremendous increase of the data volume associated to genomics [7], it could be really important to control the amount of traffic due to the transfer of genomes, reference/auxiliary files, and, for genomic cloud services, of VM images files.

Given these premises, we classify existing networking solutions for genomic services according to two main service categories: quality of service (QoS) oriented schemes, and data-driven schemes. In addition, the networking solutions applicable to large-scale geographical networks can be quite different from those deployable within datacenters. Thus, we have categorized networking solutions for the GaaS paradigm into those applicable to wide area networks (WAN), and those whose scope is limited to datacenter networks (DCN). In fact, since DCNs are usually administratively autonomous, a proprietary set of protocols and solutions may be used for managing intra-data-center traffic, leaving the standard TCP/IP stack for communications between the data center and external users [212].

### 1) Wide area network technologies for genomics

Fig. 9 shows the taxonomy of the wide area networking technology applied to Big Data in genomics. Among the approaches that we classify as QoS-oriented, we can distinguish the usage of high speed transfer protocols, and the provisioning of (virtual) link with guaranteed bandwidth. The first comment is that the two approaches are not mutually exclusive. In fact, provisioning high speed links does not guarantee that these facilities will be efficiently used by the protocol stack that manages the data transfer. For this reason, the usage of protocols able to efficiently use the available resources is important.

We can further distinguish between high speed transfer protocols that operate at layer 4 (transport protocol) and those that act at the application layer. In the first category we can consider generic, non genomic-specific transport protocol, such UDT [128], already mentioned in section VI.B since it is used in the Bionimbus PDC. It is a transport protocol based on UDP, which is able to efficiently use high speed bandwidth link, by adding reliability and congestion control to UDP in a TCP-friendly way. The availability of APIs guarantees the possibility of an easy interaction from application protocols.

As for application layer protocols, we have already mentioned that HTTP, which is commonly used to interact with public cloud to upload genomics data, is not able to provide any type of QoS. Thus, to achieve better performance, it is necessary to adopt an application protocol specifically designed to transfer large amount of data into (virtualized) computing clusters. In this regard, we can mention Globus Trasfer [214], which is the transfer service adopted by workflow-based systems using Galaxy [197]. The Globus Transfer is based on the GridFTP protocol [215] and it offers reliable, high performance, secure data transfer. Its superiority over other technologies (classic HTTP, FTP, and its variants) has been well-established when dealing with TBs of data. In addition, Globus Transfer has the capabilities to automate the task of moving files across administrative domains. In fact, it offers a "fire and forget" model in which researchers have just to submit their transfer requests, without the need to monitor the associated jobs. Globus Transfer automatically tunes the transfer parameters to maximize throughput, managing security, monitoring performance, retrying failures and recovering from faults. Finally, it notifies users of errors and job completion.

An alternative solution, used for instance by NIH [23], is the Aspera FASP protocol [222], a proprietary solution from IBM. Aspera FASP operates at the application layer, adopting UDP as transport protocol. Advantages over standard TCP are obtained by decoupling congestion and reliability control.

Due to this feature, new packets transmission is not slowed down due to the retransmission of lost packets, since retransmissions occur at the available data rate estimated inside the end-to-end path, also avoiding duplicated packets. The available bandwidth is estimated by a delay-based rate control mechanism, using measured queuing delay as the primary indication of network congestion, trying to maintain a small, stable amount of queuing in the network.

The queuing delay along the transfer path is estimated through the periodic transmission of probing packets.

A complementary approach is to provide just high-speed links/paths between facilities (as, for instance, in the ICTBioMed consortium, [23]) or VPN services with guaranteed bandwidth between data sources (e.g. research centers hosting NGS machines) and data collectors (computing clusters and/or cloud facilities), which is definitely a step forward.

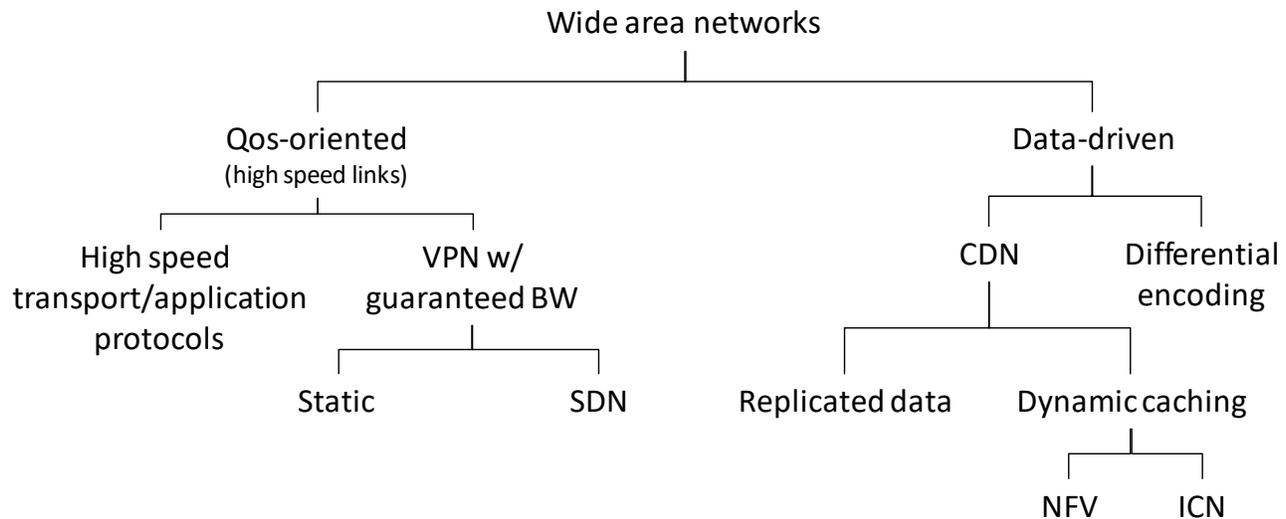

Fig. 9. Taxonomy of geographic network solutions for cloud processing of genomic data sets.

While this is a common approach in many Big Data scenarios, a static configuration does not fit well the genomics scenario. In fact, differently from applications where the number of data sources is mainly fixed (e.g. dispersed telescope sites in astronomy [107]), things are definitely more variable in genomics, where new NGS machines are continuously installed in laboratories and research centers. Thus, a static approach cannot guarantee virtual links with guaranteed bandwidth from all locations. A more interesting approach relies on novel programmable network paradigms, such as SDN and SDX [240].

As mentioned in the description of the Bionimbus PDC, it has been recently upgraded with the BioSDX facility [213], which relies on a massive usage of SDN techniques. In particular, it is able to dynamically interconnect data sources and computing nodes to support complex genomics workflows, also when these are in different network domains by means of SDX nodes. The degree of flexibility and deployability brought by SDN is really significant, since it allows to easily and incrementally integrate new sites (both sources and computing facilities), thanks to the programmability features of the SDN technology. The adoption of OpenStack for managing the cloud deployment in the Bionimbus PDC highly simplifies the adoption of SDN technologies on the network interconnecting different Bionimbus sites, thus we can globally speak of software defined infrastructure (SDI). SDI can be used to transport large genomic datasets to geographically distributed Bionimbus sites. In addition, SDI can integrate multiple and distributed (i) cloud systems, (ii) high performance (virtual) networks, and (iii) high performance storage systems. This allows supporting highly complex bioinformatics workflows, since the whole programmability of the system, from the computing to the networking part, can support it with the allocation of the (computing and networking) resources required for a specific research investigation within the Bionimbus framework.

However, as mentioned before, the allocation of networking resources to sustain high throughput transfers is not the only relevant aspect in networking when dealing with genomics. Another metric of interest is to save as much as possible network resources, mainly due to the constantly increasing volume of traffic due to networked genomics services. The networking solutions in the genomics domain aimed at savings network resources can be classified into two main tracks: CDN solutions and platforms adopting differential encoding and transfer.

As for the first approach, CDN solutions can be seen under two different flavors, that is content replication to facilitate access from distributed locations, and dynamic caching to temporarily store popular contents as close as possible to requesting users. While these solutions are often adopted in a coupled fashion, in the case of networked genomics we can bring different examples for their usage. In more detail, as mentioned before, general public clouds that offer some specific facilities to develop/run genomic pipelines, such as AWS or Google, store some very relevant contents in their storage system to speed up access to them from VMs running in their premises. In particular, AWS stores the outcome of the 1000 Genomes project [112], whereas Google allows accessing some highly relevant reference files [111] to people developing and executing genomics pipelines in its PaaS environment. Instead, as for the caching of popular contents, an interesting example is given by the already mentioned project ARES. ARES proposed a framework called Genomic Centric Networking (GCN) [177], which uses NFV caches co-located with programmable routers as general facilities to save network bandwidth, due to the increasing SDN capabilities of Géant and the facilities made available with the Géant Testbed-as-a-service (TaaS). The proposed GCN combines optimized caching supported by specifically designed signaling protocols (the OSP protocol presented in [216]), in order to minimize the overall bandwidth consumed to transfer genomes, auxiliary files, reference genomes, and VMs used to execute genomic processing, into the most convenient tenant to run the pipeline. This operation is done by leveraging cached contents to keep data transfers as local as possible. As for service time, moving contents close to the processing locations (datacenters) help to significantly reduce transfer times, which is the secondary objective of the GCN system. This system has been tested in a real prototype, consisting of a local testbed together with resources running in the TaaS. It is worth nothing that the GCN system takes some concepts from ICN [218], such as massive usage of in-network caching by means of routers empowered with cache modules, able to store content chunks. However, the genomic application scenario seems to not fit well the ICN model. In fact, the huge amount of genomic traffic call mainly for high speed, reliable, wired transfer. In such an environment, as shown in [217], standard cache modules using the HTTP protocol, such as those used in the GCN, definitely outperform ICN ones. In addition, in order to use ICN on a global scale, a naming convention has to be agreed, which seems even more difficult to achieve in the short term.

The second solution to save network bandwidth is to limit content transmission, exploiting similarities in contents to be transferred, especially for large databases of reference files. Exploiting similarity between two contents allows synchronizing large datasets with minimal consumption of network resources, since just content differences are encoded and transferred through the network. This is the case of the UDR protocol adopted in the Bionimbus PDC, which relies on the rsync Unix utility to keep synchronized some of the large genomics sets that are part of the Bionimbus Data Commons.

### 2) Datacenter network technologies for genomics

As for DCN techniques, we did not find in the relevant literature any specific solutions designed for managing datacenter traffic in genomic applications. Nevertheless, we identified a number of solutions for DCN, designed for more general classes of Big Data problems, that would fit well the requirements of genomic processing. As mentioned above, also in this case the first level classification is between QoS-oriented and data-driven approaches. We propose the taxonomy illustrated in Fig. 10.

In the first case, we can distinguish between scheduling solutions, which implements different strategies for scheduling application jobs (usually in VMs), together with their networking requirements, and solutions oriented to minimize flow completion time within DCN. For what concerns scheduling solutions, we have identified three main approaches: (i) traffic aware VM scheduling, (ii) multi-resource packing scheduling, and (iii) application-aware network scheduling.

Although there are some similarities between these approaches, they present some significant differences, which deserve a further analysis.

Traffic aware VM scheduling approaches [161] in multi-tenant datacenters were designed to handle jobs of different tenants that compete for the shared datacenter network. In order to avoid poor and unpredictable network performance, these approaches leverage explicit APIs for tenant jobs to specify and reserve virtual clusters, with both explicit VMs and network bandwidth between VMs.

Whilst most proposals include the reservation of a fixed bandwidth throughout the entire execution of a job, a nice alternative proposal is sketched in [161], where the traffic patterns of several popular cloud applications are profiled in order to find the processing phases where the network usage is significant (see also our discussion for the genomics case in section V.D). They use this information to design a fine-grained virtual network abstraction, Time-Interleaved Virtual Clusters (TIVC), that models the time-varying nature of the networking requirement of cloud applications, showing a significant increase of the utilization of the entire datacenter and a reduction in the cost of the tenants compared to the previous fixed-bandwidth abstractions.

The concept of multi-resource packing scheduling [219] consists in carrying out a joint scheduling of all resources together (CPU, memory, storage, network) when a new task is deployed. The goal of such solutions, which adopt an approach similar to multidimensional bin packing, is avoiding resource fragmentation and over-allocation of resources (especially disk I/O and network), which are typical drawbacks of current schedulers. The consequence of these policies is typically an increase of the average job completion time, with consequent waste of computing resources. In [219], the authors present Tetris, which is an implementation of a scheduler able to packs tasks to machines based on their requirements of all types of resources. Their analysis show that the designed solution can decrease the average job completion time by preferentially serving jobs that have less remaining work, without compromising too much the fairness between different jobs. Results obtained on a large cluster show an improvement of about 30% in the average job completion time while achieving nearly perfect fairness.

Finally, the third class of scheduling approach deals with scheduling of network resources by also considering the supported application. This approach is opposite to the first one (traffic aware VM scheduling). The motivation of its introduction is to comply with the common practice in cloud computing aiming to give tenant users the illusion of running their applications on *dedicated* hardware; this means that at the network level, tenant manager should have the illusion of running their application over a *dedicated* physical network, and reserve bandwidth consequently. Instead, a much more efficient approach uses a network abstraction model based on an application communication abstract structure, and not a given underlying physical network topology. The authors of [220] propose a Tenant Application Graph (TAG), a model that tenant managers can use to describe bandwidth requirements for applications. The TAG abstraction model includes the actual communication patterns of applications, thus obtaining a concise yet flexible representation.

A significant issue found when a TAG model is used to deploy VMs belonging to the same tenant is the co-location policy. In fact, in order to improve the service quality, a natural choice is to privilege co-location in the same server or rack of servers at the expense of availability of tenants. Thus, special anti-affinity policies are needed to handle this issue.

A complete different yet relevant approach is inspired by the large class of solutions aiming to minimize flow completion time (FCT). The interested reader can find a comprehensive description in [212]. The rationale behind these proposals is that flows in DCN can be mainly classified into short and long flows. Jobs that are partitioned into tasks running in multiple servers/VMs usually generate short flows (a few kB), which are associated with server requests and responses.

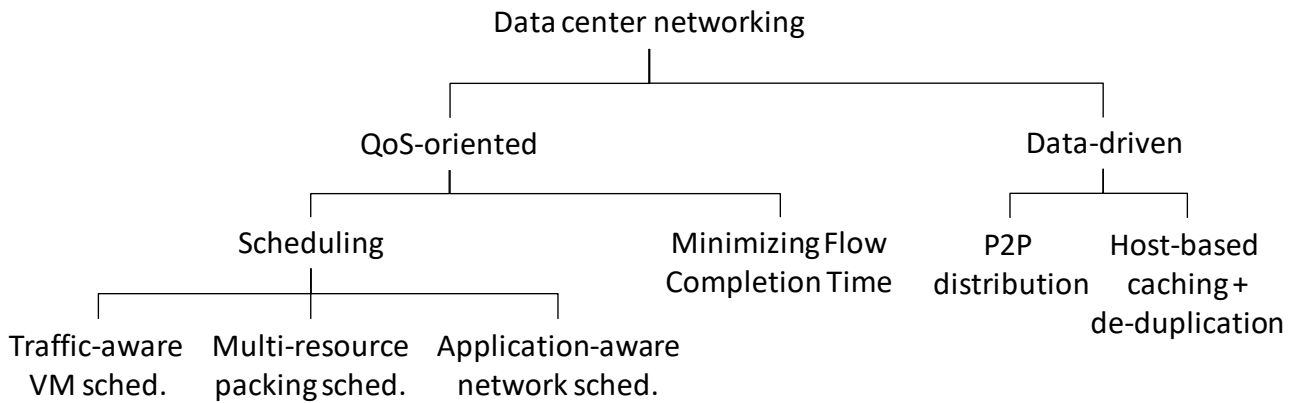

Fig. 10. Taxonomy of data center network solutions for cloud processing of genomic data sets.

Long flows may have a size of several MB or even larger, and can be used, e.g., to retrieve data from large databases. Short and long flows differ in size, nature of data, and service requirements. The first are time sensitive, whereas the latter are throughput sensitive. The requirements of these two classes of flows should be satisfied without generating conflicts with each other. In this regard, the TCP protocol seems not to be the best choice, since it tends to manage all flows equally, without considering any different features and service requirements. Many solutions have been proposed to tackle these issues. An interesting one, which seems to be suitable for genomics purposes, is CONGA [221]. It is classified as a deadline agnostic scheme, that is not aware of deadline to be met for completing flows. Basically, it is a network-based distributed congestion-aware load balancing mechanism for datacenters. CONGA exploits recent trends, including the use of multipath-forwarding capabilities in DNC. It splits TCP flows into flowlets, estimates real-time congestion on different paths, and allocates flowlets to paths based on the feedback received from remote switches. This enables CONGA to efficiently balance load and seamlessly handle asymmetry, without requiring any TCP modifications. This is an important requirement for applications designed by non-networking expert, as in the case of genomics, leaving the job of optimizing the network performance to the network itself.

Finally, the second macro-class of proposals aiming to improve efficiency in DCN is that of data-driven solutions. In this regard, we have already discussed the contrasting needs of having co-location, improving efficiency of the network resource usage (which minimizes the number of switches crossed by inter-VM traffic), and high availability in application aware-network scheduling. Instead, in this part we consider approaches designed to reduce data transmission and increase storage efficiency, with a special focus on the distribution of the VM image files. Each time a VM is started, it is necessary to load it from a dedicated local repository (e.g. Glance in OpenStack). The use of a single centralized repository could generate a bottleneck for very large settings even with high-end hardware, and could cause very large initial delay in the spawning phase in case of many concurrent requests of large images. The usage of shared storage (e.g. NAS or SAN) could eliminate image transfer, since the image catalog and nodes share the same storage backend. Thus, when an image is uploaded into the system, it becomes directly available to the physical hosts. However, this method has some drawbacks. First, as it happens in the case of genomics processing, if the application requests an intensive I/O exchange, it could underperform since many operations make use of the network. In order to avoid significant performance bottleneck, it is needed to deploy top-class hardware for both networking and storage system, thus requiring significant investments. Finally, when physical machines hosting hypervisors access shared storage, they consume resources and create undesirable *noise*. This noise was shown to have an impact on virtualized applications and is something to avoid in scientific computing environments like genomics.

Different solutions to this problem have been explored. Some proposals make use of peer-to-peer (p2p) solutions, like BitTorrent. In this way, when a new VM image file request is issued by a compute node, it is possible to take profit of the VM images cached at the physical nodes and to serve them in a distributed yet efficient fashion. In this regard, a proposal has been described in [223], where a prototype implementation has been realized on top of OpenStack. Experimental results show a significant improvement in terms of service delay with respect to a centralized distribution.

However, a key finding is that, in a real environment, the number of instances created by using the same VM image may be relatively small in a single datacenter at a given time. Thus, conventional file-based p2p distribution may not be very effective. An interesting idea consider to leverage not only caching at compute nodes (i.e. those running hypervisors), but also de-duplication techniques. This is the rationale of the paper [224], where the authors start from the known fact that different VM image files often have common chunks of data. Therefore, it results feasible and beneficial to allow chunk-level sharing among different images, thus allowing not only to save transfer bandwidth, but also increasing the number of nodes participating to the sharing process. In addition, by de-duplicating chunks with the same content, the scheme in [224] significantly saves storage space. Each host can only keep one copy of a (common) chunk in the shared cache, even when multiple VM instances make use of it. The obtained performance is really interesting, especially when the underlying physical network topology is considered in the chunk distribution. Given the possible large size of VMs hosting genomic processing software, this solution could be of great interest.

## VII. OPEN RESEARCH ISSUES

Some heterogeneous initiatives related to networked access to bioinformatics services are in progress, focusing on different issues and implementation aspects. From the analysis of the different technical approaches in the computing and networking area related to the analysis of genomic data sets, a general emerging trend for accessing these services is to make use of virtualization technologies. In particular, a massive usage of virtualization technologies is observed in both the computing (adoption of cloud computing in different flavors) and networking fields (SDN and/or NFV based solutions). However, a number of issues remain unsolved. In this section, we discuss the open issues for implementing a GaaS solution and the expected directions of the related research.

### A. Storage and retrieval issues

A peculiar aspect that is worth of further analysis is the management of cloud resources during genomic files exchange. In terms of file storage and retrieval, different issues emerge in relation to the nature of the content exchanged. For large files generated by aggregating multiple DNA reads, the NoSQL strategy seems to be quite promising, even if a substantial consensus has not been still achieved [127]. For what concerns metadata, the situation is still more uncertain since the use of proprietary processing tools based on different data structure has hindered the definition of a standardized taxonomy for effective data management [225]. Thus, an existing challenge is to achieve a global consensus in the definition of a (de facto) standard designed for both storage and data management, in order to supports data organization and retrieval independently of the software platform used. We think that a standardization effort should be pursued in order to solve this issue, and to ease the process of developing new software using standardized format for metadata.

### B. Computing and networking aspects

A number of different research challenges persists in the areas of computing and networking when dealing with genomic contents.

An important issue is data consistency when dynamic caching is used, along with the optimum joint cache management. Concerning caches, another open issue is the eviction and replacement policies of cached contents. Although the most frequently used policy to evict contents from a cache, upon storage space exhaustion, is the least recently used (LRU) [145], in case of genomic contents LRU could not be the optimal choice. Future research is expected to investigate how to optimally decide if a genomic content should be evicted, not only based on its estimated popularity, but also of the content distribution in other nearby caches, in conjunction with de-duplication techniques [155].

A further open issue is the optimal content placement in the available caches to build an efficient GaaS system. Whereas the information centric networking (ICN) community has already addressed this issue (see, for instance [158]), it is completely open for genomic contents, which can exhibit unexpected relationships across different analyses [159][160]. It is our opinion that a measurement-assisted probabilistic criterion, similar to that used in [158], could be sub-optimal. However, for both content eviction and content placement, it is necessary to further investigate the request patterns of the genomic services already made available in clouds in order to extract evidence from real data and design optimal policies.

Another issue is related to the adoption of advanced scheduling of service requests in data centers, such as those discussed in section VI.C.2. Currently, network services are characterized by known traffic profiles, and this information can be taken into account when the instantiation of VMs running these services are scheduled [161]. However, profiling network demand for genomic services is still early, beyond some examples like those reported in section V.D. For instance, in that experiment, which uses a single VM for running the genome processing pipeline, the network is used only during the initial upload of the VM image, and the subsequent upload of auxiliary/reference files and of the genomic data set to process. Then, the network remains inactive until the pipeline generates a result and transfers it to the database. However, this last step can be neglected, since the size of the output file is usually much smaller than that of input files. Nevertheless, with the genome processing evolving in distributed and more complex processing pipeline over more than a VM, this scenario could radically change, and the traffic profile should be accounted in order to optimize a GaaS system.

Finally, another open issue is the selection of an efficient network solution for upload genomic data sets into the cloud, or between datacenters [9]. Candidates are application layers techniques, such as the Aspera solution [222], the UDT protocol used in Bionimbus, built on top of the UDP protocol [128], or TCP extensions for high speed networking [162].

### C. Exploiting intrinsic features of genomic contents

Although initiatives aiming at optimizing the networked genomic services exist, up to the knowledge of the authors, no significant results still exist for exploiting the intrinsic features of genomic contents and their specific usage. For example, it is well known that different macroscopic evidences of the genetic contents, commonly referred to as genotypes, share a substantial portion of the human genome. This commonality could be used for further optimizing both the medical practices (which is definitely out of the scope of this paper) and the file management. The different needs and specifications of genomic services, can be translated into different optimization problems, with different cost functions reflecting the most important performance figures to be optimized, such as the service delivery time, the operating costs, the efficiency of network protocols, or the number of service types and the total number of services being delivered simultaneously.

### D. Improvements in bioinformatics software packages

A further research challenge, currently investigated by some initiatives, aim to overcome some structural limitations of the most popular bioinformatics software packages through parallel programming and distributed processing. In particular, from one side we expect that large genomic datasets will be processed through MapReduce-based services, if the required software is available. However, when the dataset is limited, or the software has not been ported into the MapReduce paradigm, the current implementation of bioinformatic software exhibits significant limitations in exploiting the parallelism made available by multi-core computing architecture. Open issues are related to the need of an almost complete refactoring of the pipelines, to exploit parallelism in multi-core computing architectures. Also the usage of FPGA-based hardware accelerators, combined with MapReduce frameworks, opens new scenarios for efficiently handling genomic data sets of modest size.

### E. Security and privacy issues

A lot of issues in the field of security and privacy are arising, related to the increasing diffusion of high throughput genetic sequencing technologies in parallel to the ubiquitous Internet. For instance, direct-to-consumer personal genomic information (*personalized genomics*) is a hot topic, including significant issues such as authorization and authentication services, ethical issues in managing genomic files, privacy and security [17][174][228].

In the era of Big Data, the capability to protect medical and genomics data is a challenging issue [163][164]. In fact, although cloud computing can provide massive processing power, often in the life science sector it is perceived as insecure. Although it seems more like a prejudice, being in many cases cloud solutions as much secure, if not better, than local security policy, clinical sequencing must meet regulatory requirements and cloud computing has to be cautiously considered [9]. However, given the essential benefits achievable by cloud processing of genomic data sets, its adoption is out of question, and a comprehensive legislation in this area is expected [165].

A special mention is due to the encryption mechanisms, the vulnerabilities of interfaces for network-based from customers, the replication techniques used in case of disaster recovery, and the data deletion techniques for the reassignment of virtual resources [9].

In addition, the continuous progresses in the field of advanced decryption and de-anonymisation techniques raise serious challenges for the guarantee on anonymity when using "anonymised" sequenced data, especially when these data are combined with other public sources and Internet searches [227]. In the case of highly sensitive data, as it often happens, when all the recommended practices are used, the weakest link is the human operator. Although up to now most re-identification attempts have been carried out by academic experts, there are concerns about privacy leaks if large-scale hacking results about de-identified genomic data would be available to the general public. In [226], it is suggested that some specific algorithmic methods, such as partial homomorphic encryption, secure multiparty computation or differential privacy, could provide the necessary privacy-protecting layer for genomic data. However, it is still to be investigated the effort to integrate these algorithms into bioinformatics pipelines and tools. Clearly, only the availability of a secure and privacy-preserving platform will encourage the flowing of genomic data across silos and reduce complexities around patient informed consent procedures. In this regard, the paper [226] includes an interesting analysis of the security standards to be adopted by cloud providers service providers (e.g. DNANexus [196]). Usually, IaaS providers are required to implement security policies at the cloud infrastructure layer, whereas the service provider has to secure the interface made available to the final users.

Data tenancy is a further challenge. It is associated with the availability of data stored within a commercial cloud in the case the cloud service provider should dismiss the service. Open issues are related to need of quickly moving huge volume of data to another provider, and on the interoperability among the two players, which could make this transition tricky to manage.

## VIII. Conclusion

This paper illustrates a big picture of the implementation of genomic processing services in modern infrastructure settings, with a special focus on the related computing and networking issues. To this aim, we introduce the concept of GaaS, which accounts for both networking and computing solutions specifically designed for handling genomic data sets. This research area has recently gained a lot of momentum due to the significant decrease of sequencing costs, which has produced an inversely correlated increase of bioinformatics services and relevant data.

In this paper, we have highlighted the peculiar features of genomic data and services, and the expected impact on network protocols in both service and control planes. We have highlighted that, although the evolutions of the sequencing technologies has increased both the genomic data generation rate and the number of genomic services, the support for the computing and networking side is still limited. What has emerged is that a lot of research is still needed, although some research initiatives are already focused on the most relevant challenges. In particular, the common features of these initiatives is the usage of the cloud infrastructures (either public or private) to offer genomic processing services. In addition to the computing services, some of these projects have started addressing also the related networking aspects, due to the huge amount of networking resources needed to handle genomic data sets in the cloud. Some of them have already included in the computing platform specific reference data, to limit the amount of required traffic, whereas other try to optimize the transfer by using high speed protocols and/or caching techniques.

We have also discussed a number of open issues for building a GaaS system, which are particularly relevant in the fields of storage, computing, networking, and security. Finally, we have also emphasized that, since genomic data sets are very rich sources of information, novel techniques should be developed for exploiting mutual information between different analyses. For all these reasons, it is expected that this research area will be very extremely productive over the next few years.

## Acknowledgement

The authors would like to thank the anonymous reviewers for their valuable comments and suggestions to improve the quality of the paper.

This work has been supported by the EU Horizon 2020 project 5G-EVE (grant No. 815974), and by projects CLOUD and HYDRA funded by the University of Perugia.